\documentclass{aa}
\usepackage{graphicx}
\usepackage{txfonts}
\usepackage{lipsum}
\usepackage{subcaption}
\usepackage{lscape}
\usepackage{placeins}
\usepackage{natbib}
\usepackage{amsmath}
\begin{document}

\title{Polarization of GRB standard X-ray afterglow and its detection prospects by eXTP}

\author{Jun-Xi Lin\inst{1}
\and Zelin Ren\inst{1}
\and Mi-Xiang Lan\inst{1}\fnmsep\thanks{Corresponding author: lanmixiang@jlu.edu.cn}
\and Ming-Yu Ge\inst{2}
\and Zi-Gao Dai\inst{3}
\and Xue-Feng Wu\inst{4}
\and Shuang-Nan Zhang\inst{2, 5}
}
\institute{Center for Theoretical Physics and College of Physics, Jilin University, Changchun, 130012, China; lanmixiang@jlu.edu.cn
\and Key Laboratory of Particle Astrophysics, Institute of High Energy Physics, Chinese Academy of Sciences, Beijing, 100049, China
\and Department of Astronomy, School of Physical Sciences, University of Science and Technology of China, Hefei, 230026, China
\and Purple Mountain Observatory, Chinese Academy of Sciences, Nanjing, 210008, China
\and University of Chinese Academy of Sciences, Beijing, 100049, China
}

\date{Received September 30, 20XX}

\abstract
{The polarization signatures of Gamma-ray Burst (GRB) afterglows serve as a powerful diagnostic tool for studying their environments and jet physics.}
{This work systematically investigates the X-ray (2--8~keV) polarization properties of standard GRB afterglows and assesses their detectability with the Polarimetry Focusing Array aboard the enhanced X-ray Timing and Polarimetry (eXTP) satellite.}
{A Morris global sensitivity analysis is first conducted to identify the dominant parameters, which are then assigned observationally motivated probability distributions. In particular, the isotropic energy, half-opening angle, and initial Lorentz factor are sampled jointly via a Gaussian copula to reproduce the empirical Ghirlanda and Liang correlations. Monte Carlo simulations of $10^{3}$ afterglows are performed and validated against the observed 10~keV flux distributions of a selected Fermi--Swift sample (K--S $p = 0.29$ at $10^{3}~\mathrm{s}$ and $p = 0.18$ at $10^{4}~\mathrm{s}$).}
{The simulations yield an overall polarization event rate of $\lesssim 1.5\%$ for standard GRB X-ray afterglows with eXTP/PFA, reflecting the intrinsically low polarization produced by a random magnetic field confined to the shock plane. The optimal detection window occurs near the jet break at late times, when the PD peaks. For exceptionally luminous events such as GRB~221009A, however, the PD remains above the MDP over the full interval $10^{3}$--$10^{6}~\mathrm{s}$, demonstrating that eXTP/PFA can capture nearly the entire polarization evolution for such rare, bright bursts.}
{}

\maketitle
\nolinenumbers

\section{Introduction}
Gamma-ray bursts (GRBs) are extreme transient events characterized by an intense initial flash of gamma-rays followed by multi-wavelength afterglows.
GRBs are primarily classified by their $T_{90}$ duration (the time range accumulating $90\%$ of the total gamma-ray flux) into long-duration GRBs (lGRBs, $T_{90} > 2$ s) and short-duration GRBs (sGRBs, $T_{90} < 2$ s) \citep{1993ApJ...413L.101K}, with rare exceptions such as GRB 211211A \citep{2024ApJ...976...62Z}.
lGRBs typically originate from the collapse of massive stars \citep{1986ApJ...308L..43P, 1993ApJ...405..273W, 2003ApJ...599L..95M, 2007ApJ...655L..25Z}, while sGRBs are generally associated with the merger of compact objects \citep{2017ApJ...848L..13A, 2017ApJ...848L..14G, 2018PhRvL.120x1103L}.
The polarization signatures of GRB afterglows provide a critical probe of the magnetic field configurations (MFCs) in the radiation region.
In the forward shock region, turbulent fields generated by Weibel instability \citep{1959PhRvL...2...83W, 1999ApJ...526..697M, 2015NatPh..11..173H} lead to low polarization; in contrast, the reverse shock region may preserve large-scale ordered MFCs advected from the central engine, resulting in higher polarization \citep{2009Natur.462..767S, 2013Natur.504..119M}.
For black hole central engines, the Blandford-Znajek mechanism produces toroidal-dominated fields at large radii \citep{1977MNRAS.179..433B}, while for a magnetar central engine, the MFC may be coherently aligned with each other on large scales \citep{2001A&A...369..694S}.
\par The circumburst environment further modulates polarization signals. Fits to observations indicate a stratified density profile $n(r) \propto r^{-1}$ \citep{2013ApJ...776..120Y} rather than a uniform interstellar medium (ISM) with constant density or a stellar wind with number density scaling as $n(r) \propto r^{-2}$.
Such gradients, combined with the Equal-Arrival-Time Surface (EATS) effect \citep{1998ApJ...497L..17S, 2000ApJ...543...90H}, introduce a complex time dependence of the polarization in GRB afterglows \citep{2023ApJ...952...31L}.

\par Testing these theoretical predictions requires sensitive X-ray polarimetric instrumentation. The enhanced X-ray Timing and Polarimetry mission (eXTP) is a major international space science project. It aims to study the laws of physics under extreme gravity, magnetic fields, and density by observing phenomena such as black hole event horizons and the extreme environments of neutron stars \citep{2025SCPMA..6819502Z, 2025SCPMA..6819503L, 2025SCPMA..6819504B, 2025SCPMA..6819505G}. Aboard the satellite, the Polarimetry Focusing Array (PFA) is designed to conduct polarimetry specifically in the 2--8 keV X-ray band \citep{2022ApJ...934..109Q, 2023ExA....56..517Q, 2025SCPMA..6819502Z}.

\par The standard model generally provides a consistent description of the optical afterglow. In the X-ray band, however, significant deviations are frequently observed, including shallow-decay plateaus and X-ray flares, which violate the standard closure relations \citep{2006ApJ...642..354Z, 2006ApJ...642..389N, 2007ApJ...662.1093W}. 
These features suggest the presence of additional emission components or late-time energy injection. The polarization properties of these additional X-ray components have been investigated in several targeted studies: for X-ray flares, time-integrated polarization is predicted to be very high for large-scale ordered magnetic fields and near zero for random magnetic fields \citep{2018ApJ...862..115G, 2025ApJ...989..172W}; for the shallow-decay plateau phase, the relativistic wind bubble model predicts a polarization well above the detection threshold of eXTP/PFA, whereas the structured ejecta model predicts one generally below the threshold \citep{2016ApJ...816...73L, 2026ApJ...996...33W}. 
In this context, the predictions presented in this work — based on the standard external-shock afterglow model with a two-dimensional random magnetic field — provide a baseline for the standard afterglow segments. Any observed departures from these baseline predictions, combined with the distinct polarization signatures expected from flares and plateaus, would offer a comprehensive diagnostic of the central engine activity across different afterglow stages, making eXTP/PFA a powerful tool for probing the full diversity of GRB X-ray afterglows.

\par Despite extensive theoretical studies \citep{1999ApJ...520..641S, 1999MNRAS.309L...7G, 2003Natur.423..388W, 2004MNRAS.354...86R, 2010xpnw.book..202L, 2016ApJ...816...73L, 2023ApJ...952...31L, 2024NatAs...8..134A}, none of these studies were focused to predict the detection prospect of the upcoming eXTP mission \citep{2025SCPMA..6819502Z, 2025SCPMA..6819506Y}. 
In this paper, we systematically investigate the polarization properties of GRB afterglows predicted by the standard external-shock model in the X-ray band (2--8 keV). A Morris global sensitivity analysis is conducted in Section \ref{sec:3} to rank the model parameters by their influence on the X-ray polarization and to discard those with negligible impact. 
The retained key parameters are assigned statistical distributions derived from observations and fed into Monte Carlo simulations to assess the polarization detection prospects with eXTP/PFA, yielding an expected event rate of ($\leq1.5\%$), as presented in Section \ref{sec:4}.
We apply our model to interpret the observations of GRB 221009A and investigate the detection ability of such a bright event by eXTP/PFA in Section \ref{sec:5}.
Finally, our conclusions and discussion are presented in Section \ref{sec:6}.

\section{The model}  
\label{sec:2} 
The afterglow is widely described by the forward and reverse shock model \citep{1999ApJ...520..641S,2016ApJ...816...73L}.
The interaction structure comprises four radially stratified regions from the innermost to outermost layer: (4) unshocked fireball material; (3) shocked fireball material; (2) shocked ISM or circumstellar wind material; (1) unshocked ISM or circumstellar wind material.
The dynamic model here follows that in \cite{2016ApJ...816...73L}, which fully accounts for the effects of the reverse shock, with the assumption of adiabatic shocks.
We consider a stratified medium with a power-law density profile \citep{2013ApJ...776..120Y}:
\begin{equation}
n(r) = n_0 \left( \frac{r}{r_0} \right)^{-k},
\end{equation}
where $r$ denotes the radial distance from the central engine and $n_0 = 1~\text{cm}^{-3}$. The relativistic outflow is simulated using a top-hat jet configuration with half-opening angle $\theta_j$, neglecting the lateral expansion \citep{2012MNRAS.421..570G}. We account for the EATS effect, where photons emitted from a radius $r$ at source-frame time $t$ reach the observer at an observed time $t_{\rm obs}$ given by \citep{1998ApJ...497L..17S}:
\begin{equation}
t_{obs} = ({t - \frac{r \cos\theta}{c}})(1+z),
\end{equation}
where $\theta$ is the angle between the velocity of the jet element and the line of sight, $c$ is the speed of light, $z$ is the redshift of the source, and we assume constant initial Lorentz factor $\eta$ prior to reaching the characteristic emission radius $r_b$.

The wide energy band and wide field of view detector W2C onboard eXTP can be self-triggered on orbit. Since the pointing velocity is 18 degree per minute, the PFA can point to the target with 10 seconds fastest response time and with 200 seconds ordinary response time, which makes the detection of the early GRB X-ray afterglow polarization possible. However, multi-wavelength spectral analysis reveals that for the GRB afterglows where a reverse shock component has been detected in the optical and radio bands, no corresponding reverse shock emission is observed in the X-ray band \citep{2019ApJ...878L..26L, 2019ApJ...884..121L, 2018ApJ...862...94L, 2018ApJ...859..134L, 2016ApJ...833...88L, 2013ApJ...776..119L}. Additionally, the calculated X-ray flux ratio $f_{\nu,3}/f_{\nu,2} \ll 1$ (The $f_{\nu,3}$ and $f_{\nu,2}$ are the flux density of the reverse-shock and forward-shock regions, respectively.), this is because the characteristic synchrotron frequency of reverse shock emission lies far below the X-ray observing band, rendering its contribution to the total X-ray flux negligible. Thus, forward shock emission dominates the entire standard X-ray afterglow. We assume the forward shock generates a two-dimensional random magnetic field confined to the shock plane. All primed quantities refer to the comoving frame of the outflow. For a single electron with Lorentz factor $\gamma_e$, the spectral power at frequency $\nu'$ is:
\begin{equation}
p'(\nu') = \frac{\sqrt{3}e^3 \sin{\theta_B'}B'}{m_e c^2} F\left( \frac{\nu'}{\nu'_c} \right),
\label{eq:single_e}
\end{equation}
where $\nu'_c = \gamma_e^2 e \sin{\theta_B'}B'/ 2\pi m_e c$ is the characteristic frequency, and $\theta_B'$ is the pitch angle of the electrons in the magnetic field \citep{1979rpa..book.....R}.
The comoving frequency is related to the observer's frequency $\nu$ by the Doppler factor $\mathcal{D}$ and redshift $z$:
$\nu^{\prime} = \nu (1 + z)/\mathcal{D} \quad ,\mathcal{D} = 1/\Gamma \left(1 - \beta_{V} \cos \theta \right)$.
$\beta_{V}$ is the dimensionless velocity of the shell and $\Gamma$ is the bulk Lorentz factor.
Given the electron energy distribution ($n(\gamma_e)\propto \gamma_e^{-p}$), the radiation power is:
\begin{equation}
P'(\nu') = \int p'(\nu') n(\gamma_e) d\gamma_e,
\end{equation}
The flux density $f_{\nu,2}$ and Stokes parameters for linear polarization ($Q_{\nu,2}$) can be found in \cite{2023ApJ...952...31L}.
The polarization degree (PD) of the emission from the forward shock region can be calculated by:
\begin{equation}
PD = \frac{Q_{\nu,2}}{f_{\nu,2}},
\end{equation}

\section{Polarization properties and parameter sensitivity}
\label{sec:3}

The standard external-shock model involves a number of free parameters. To assess the detection prospects of eXTP/PFA in Section~\ref{sec:4}, we must identify which of these parameters dominate the predicted X-ray polarization and which contribute only marginally. In this section, we first summarize the key polarization features already established for the standard afterglow model (Section~\ref{sec:3.1}), and then perform a Morris global sensitivity analysis to quantitatively rank the parameters and screen out the negligible ones (Section~\ref{sec:3.2}).

\subsection{X-ray polarization}
\label{sec:3.1}
For a uniform top-hat jet with synchrotron emission from a random magnetic field confined to the shock plane, the PD is tightly coupled to the light-curve evolution. Two PD peaks are expected: the first when $1/\Gamma \sim \theta_{j} - \theta_{v}$, and the second when $1/\Gamma \sim \theta_{j} + \theta_{v}$, the latter roughly coinciding with the jet break time \citep{1999ApJ...524L..43S, 1999MNRAS.309L...7G, 2004MNRAS.354...86R, 2010xpnw.book..202L, 2023ApJ...952...31L}. Because the polarization is geometric in origin, it depends only weakly on photon frequency above the optical band \citep{2004MNRAS.354...86R, 2010xpnw.book..202L}.
\citet{2013ApJ...776..120Y} introduced a stratified circumburst medium $n(r) \propto r^{-k}$ and fitted its parameters to observational data. Adopting this density profile, \citet{2023ApJ...952...31L} studied the effects of the viewing geometry ($q \equiv \theta_{v}/\theta_{j}$), the density index $k$, and the characteristic radius $r_0$ on the afterglow polarization. However, because these studies varied parameters individually, the relative importance of the different parameters remains unknown. A global sensitivity analysis is required, which we perform in Section~\ref{sec:3.2}.

\subsection{Morris global sensitivity analysis}
\label{sec:3.2}
\subsubsection{Method and parameter setup}
\label{sec:3.2.1}

To identify which parameters dominate the predicted X-ray polarization and which can be safely fixed, we adopt the Morris global sensitivity analysis \citep{Morris01051991}, as refined by \citet{CAMPOLONGO20071509}. For a model with $K$ input parameters, the method generates $R$ trajectories through the input space. Each trajectory consists of $K+1$ points obtained by perturbing one parameter at a time by a fixed step $\Delta$, thus producing one elementary effect per input per trajectory. The elementary effect of the $i$-th parameter at a given point $\mathbf{X}$ in the normalized input space ($[0,1]^K$) is defined as
\begin{equation}
d_i(\mathbf{X}) = 
\frac{y(X_1,\dots,X_i+\Delta,\dots,X_K) - y(\mathbf{X})}{\Delta},
\end{equation}
where $y$ is the model output of interest. Following the standard prescription, the number of grid levels is chosen as $p = 6$ (even), and the step size is set to $\Delta = p/[2(p-1)] = 0.6$. From the $R$ elementary effects sampled for each parameter, two statistics are derived. The first,
\begin{equation}
\mu_i^* = \frac{1}{R}\sum_{j=1}^{R} \bigl|d_i^{(j)}\bigr|,
\label{eq:mu_star}
\end{equation}
is the mean of the absolute elementary effects. The use of absolute values, introduced by \citet{CAMPOLONGO20071509}, 
avoids the cancellation of elementary effects of opposite sign that can occur when the model response is non-monotonic. The second,
\begin{equation}
\sigma_i = \sqrt{\frac{1}{R-1}\sum_{j=1}^{R} 
\bigl(d_i^{(j)} - \bar{d}_i\bigr)^2},
\qquad \bar{d}_i = \frac{1}{R}\sum_{j=1}^{R} d_i^{(j)},
\label{eq:sigma}
\end{equation}
is the standard deviation of the elementary effects. A large $\sigma_i$ indicates that the parameter's influence varies strongly across the input space, either because the dependence is nonlinear or because the parameter interacts with others. In the $\mu^*$--$\sigma$ plane, parameters with both small $\mu_i^*$ and small $\sigma_i$ (i.e., those clustering near the origin) have negligible influence and are candidates for fixing. 
To facilitate a unified ranking, we also introduce a composite index
\begin{equation}
B_i = \sqrt{\mu_i^{*2} + \sigma_i^2},
\label{eq:B_index}
\end{equation}
which treats the overall influence and the nonlinearity or interaction effects on an equal footing.
We consider $K = 11$ model parameters, whose physical ranges are listed in Table~\ref{tab:morris_param}. Several choices are made to reduce the effective degrees of freedom without loss of generality. First, the density profile $n(r) = n_0 (r/r_0)^{-k}$ contains a degeneracy between $n_0$ and $r_0$: any rescaling of $n_0$ can be exactly compensated by a corresponding shift in $r_0$. We therefore fix $n_0 = 1~\mathrm{cm}^{-3}$ and treat $r_0$ as a free parameter, which spans the full range of physically relevant density normalizations. Second, the observer's viewing angle $\theta_v$ is not treated as an independent parameter. Instead, we fix the ratio $q \equiv \theta_v / \theta_j = 2/3$ \citep{1999MNRAS.309L...7G}. This choice is motivated by the well-established result that $q$ strongly influences the predicted polarization. Fixing $q$ to the canonical value allows us to focus the sensitivity budget on the remaining eleven parameters. For the Monte Carlo simulations of Section~\ref{sec:4}, however, $q$ is naturally treated as a free parameter and sampled from its full distribution.
The parameters are sampled independently from uniform distributions within their specified ranges. Parameters spanning several orders of magnitude are sampled uniformly on a logarithmic scale to ensure adequate coverage of both small and large values; the sampling scheme adopted for each parameter is given in the final column of Table~\ref{tab:morris_param}.
The six quantities considered are the X-ray flux density and the PD at 10~keV, each evaluated at three representative observer times: $t_{obs} = 10^3$~s, $10^4$~s, and $10^5$~s. We denote the flux densities by $f_3$, $f_4$, $f_5$ and the PD by ${PD}_3$, ${PD}_4$, ${PD}_5$ at those three times, respectively. The sensitivity ranking obtained at 10~keV applies equally to the 2--8~keV eXTP/PFA band.
We adopt $R = 150$ trajectories, each consisting of $K+1 = 12$ model evaluations, giving a total of $R(K+1) = 1800$ model evaluations. The trajectories are generated in the normalized input space $[0,1]^K$ and then mapped to the physical parameter ranges via the transformations specified in Table~\ref{tab:morris_param}. The corresponding model outputs are computed by evaluating the afterglow model at each node of the design matrix.

\begin{table}[t]
\centering
\caption{Parameters of the Morris sensitivity analysis and their sampling ranges.}
\label{tab:morris_param}
\begin{tabular}{lcccc}
\hline
Parameter          & Symbol                              & Range                & Sampling   \\
\hline
Density index       & $k$                                & $[0,\, 2]$           & lin       \\
Characteristic radius & $r_0$                            & $[15,\, 18]$         & $\log_{10}$       \\
Isotropic energy    & $E_{iso}$                          & $[50,\, 54]$         & $\log_{10}$       \\
Initial Lorentz factor & $\eta$                          & $[2,\, 3]$           & $\log_{10}$       \\
Initial outflow width       & $\Delta_0$                 & $[10,\, 13]$         & $\log_{10}$       \\
Emission radius     & $r_{b}$                            & $[13,\, 15]$         & $\log_{10}$       \\
Half-opening angle   & $\theta_{j}$                      & $[-4.5,\, -1]$       & $\ln$        \\
Electron fraction   & $\epsilon_{e}$                     & $[-2.0,\, -0.5]$     & $\log_{10}$       \\
Magnetic fraction   & $\epsilon_{B}$                     & $[-4.5,\, -1.5]$     & $\log_{10}$       \\
Electron index      & $p$                                & $[2.0,\, 3.0]$       & lin       \\
Redshift            & $z$                                & $[0.2,\, 5.0]$       & lin       \\
\hline
\end{tabular}
\tablefoot{The final column indicates the sampling method: ``lin'' = uniform in linear space; ``$\log_{10}$'' = uniform in $\log_{10}$; ``ln'' = uniform in $\ln$. All parameters are sampled independently.}
\end{table}

\subsubsection{Screening results}
\label{sec:3.2.2}

\begin{figure*}[ht!]
\centering
\includegraphics[width=0.48\linewidth]{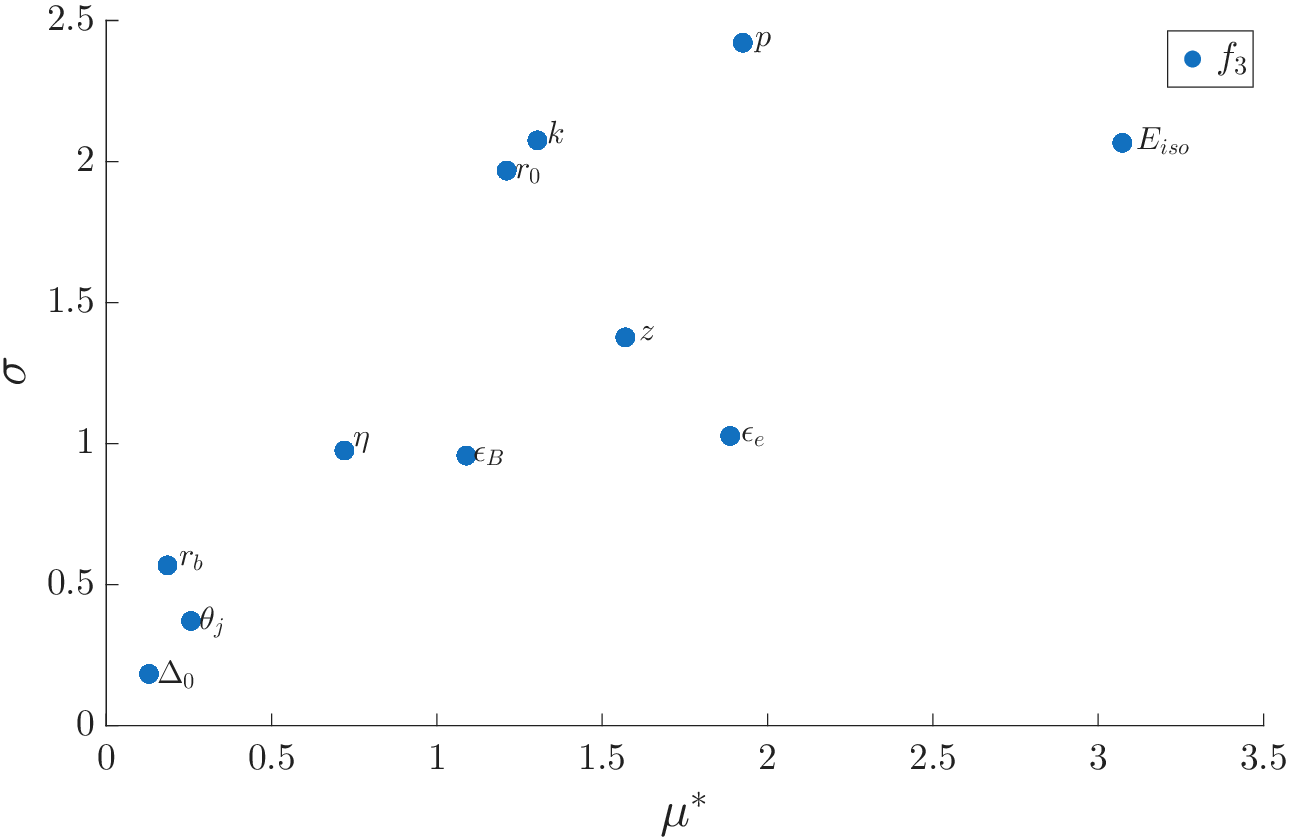}
\includegraphics[width=0.48\linewidth]{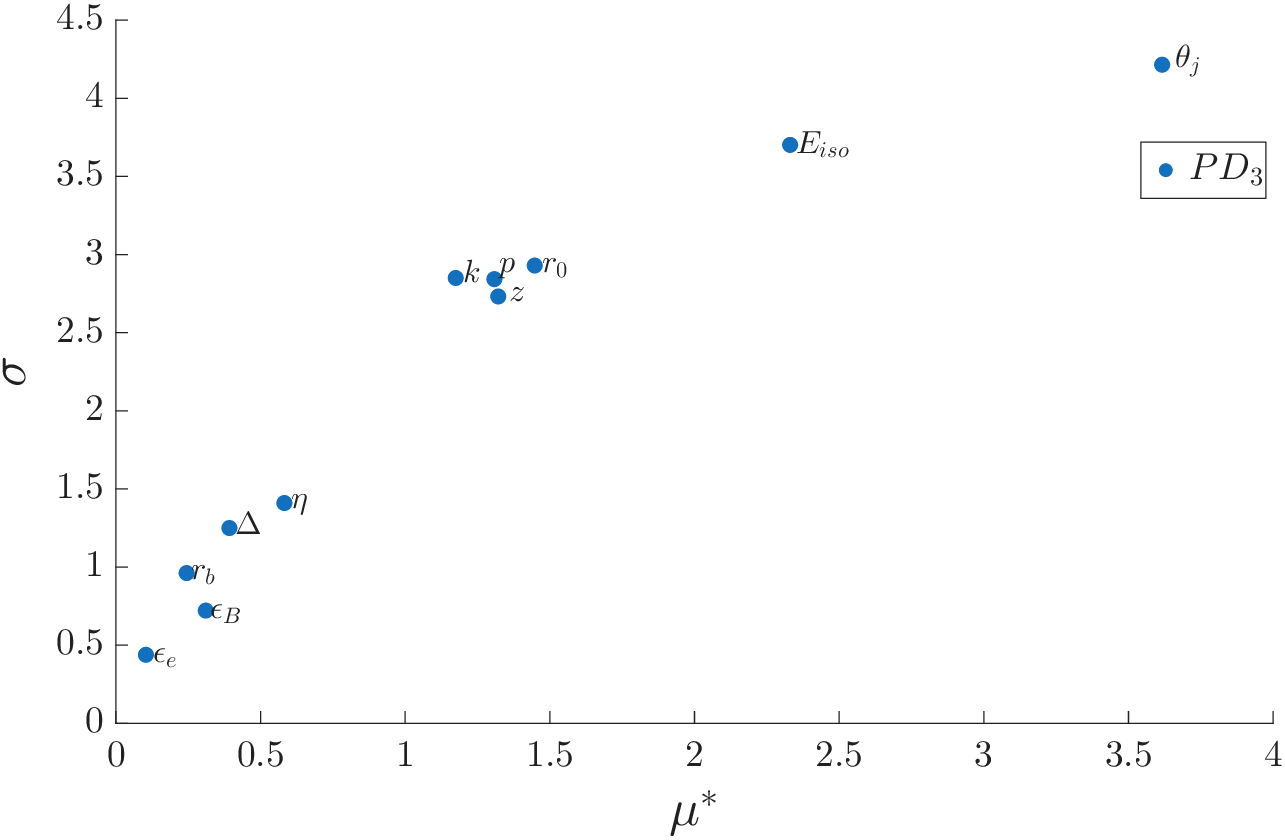}
\medskip
\includegraphics[width=0.48\linewidth]{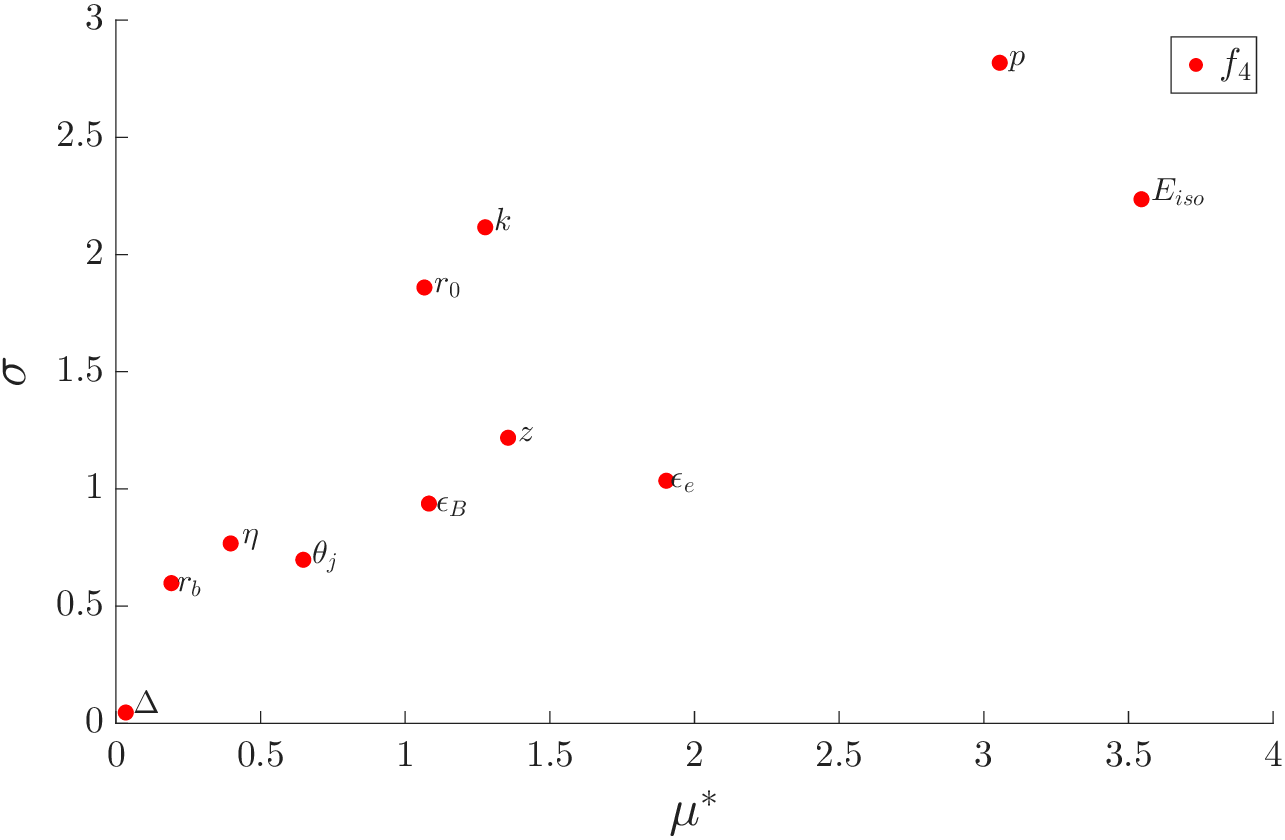}
\includegraphics[width=0.48\linewidth]{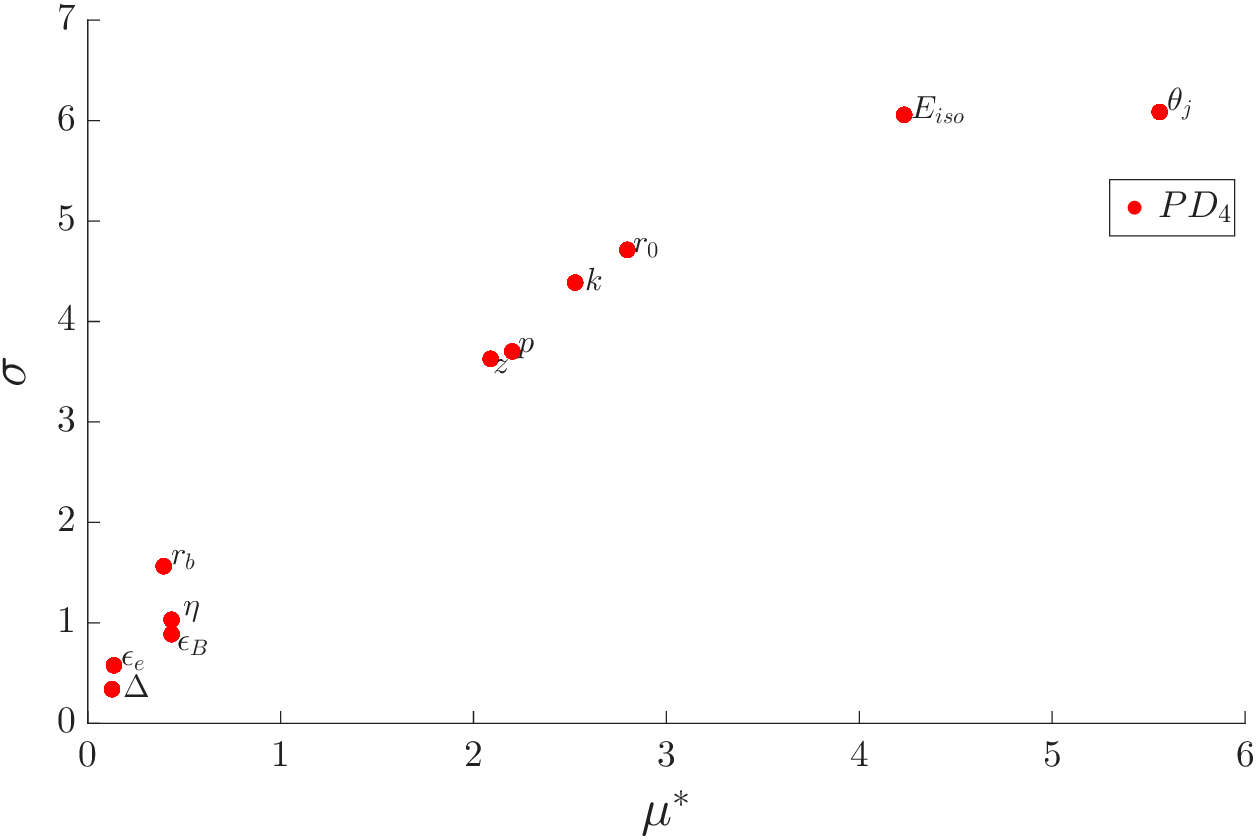}
\medskip
\includegraphics[width=0.48\linewidth]{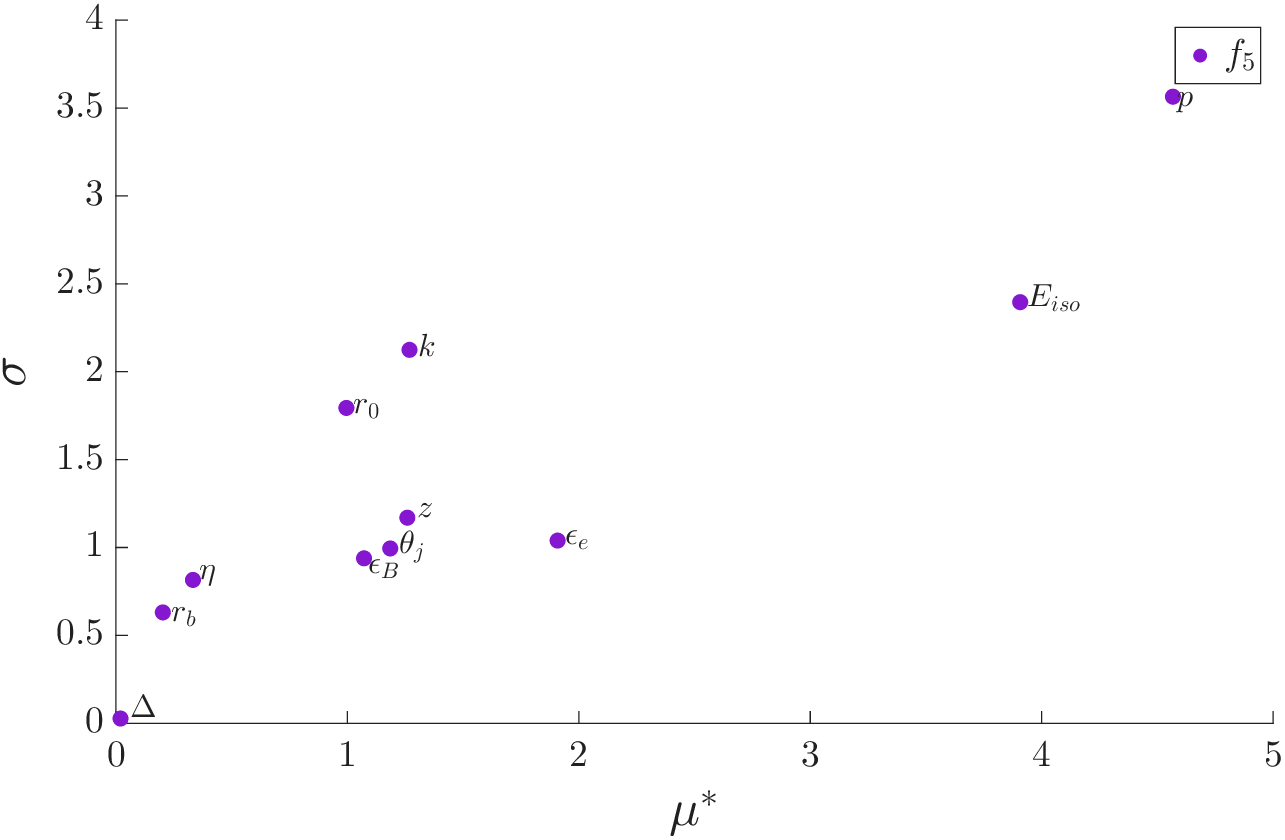}
\includegraphics[width=0.48\linewidth]{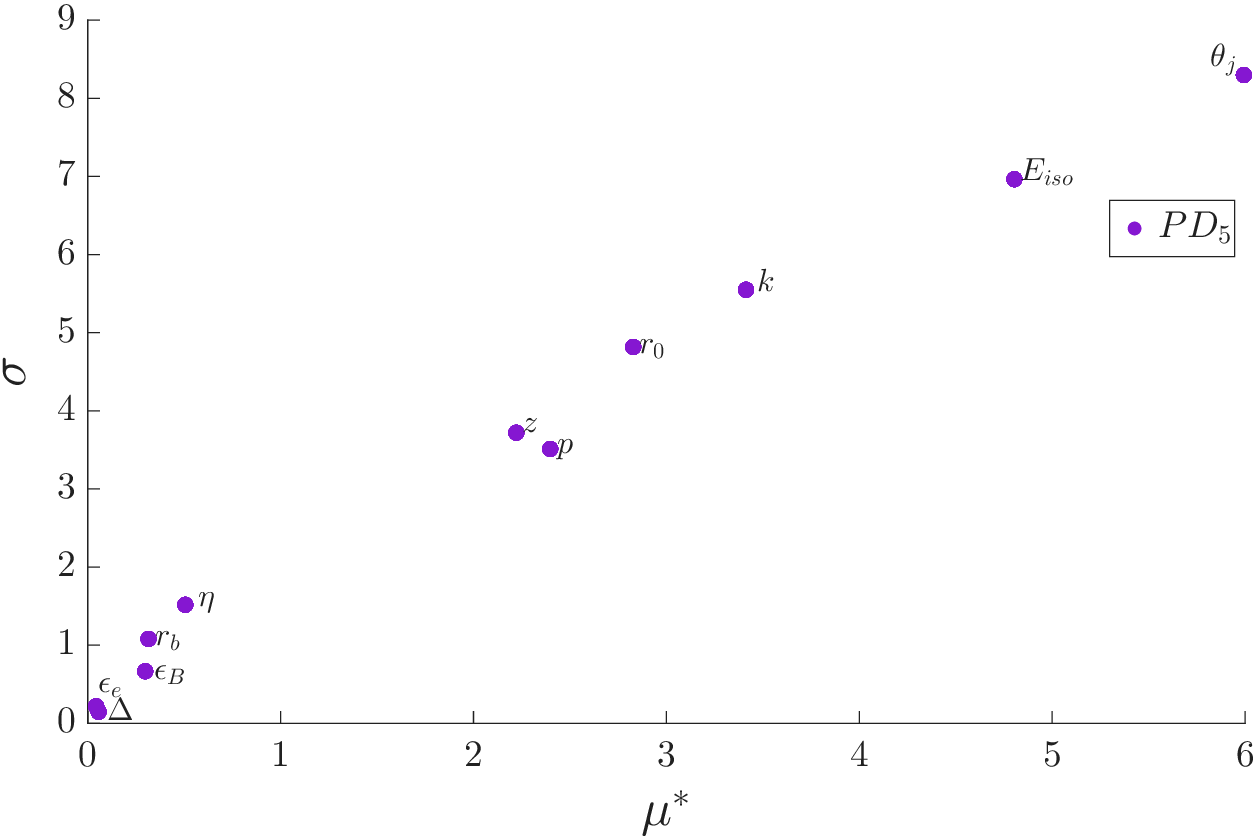}
\medskip
\caption{Morris global sensitivity analysis: $\mu^*$--$\sigma$ scatter plots for eleven model parameters. 
\textit{Left column:} flux density at 10~keV at $t_{ obs} = 10^3$~s ($f_3$), $10^4$~s ($f_4$), and $10^5$~s ($f_5$).
\textit{Right column:} corresponding PD (${PD}_3$, ${PD}_4$, ${PD}_5$).}
\label{fig:1}
\end{figure*}

Fig.~\ref{fig:1} shows the $\mu^*$--$\sigma$ scatter plots for eleven parameters evaluated on the six target quantities ($f_3$, $f_4$, $f_5$, ${PD}_3$, ${PD}_4$, ${PD}_5$). Parameters clustering near the origin have negligible influence and are candidates for fixing; those in the upper-right region dominate the model response and/or exhibit strong nonlinearity or interactions. The most prominent result is the separation between the parameters that control the flux and those that control the PD.

Flux sensitivity:
The X-ray flux density is primarily driven by $E_{iso}$ and $p$, the latter becoming increasingly important at later times as the spectral slope governs the late-time flux decay. The parameters $\epsilon_{e}$, $\epsilon_{B}$, $z$, $k$ and $r_0$ all exert a moderate influence on the flux, $\theta_{j}$ and $\eta$ play only a minor role, although the influence of $\theta_{j}$ grows at later times as the jet break is approached. $\Delta_0$ and $r_{b}$ are negligible for the flux across all times.

Polarization sensitivity:
For the PD, the parameter ranking differs markedly from that of the flux. $\theta_{j}$ and $E_{iso}$ dominate the PD, with large interaction effects that reflect the behavior of the dynamics ($1/\Gamma \sim \theta_{j} \pm \theta_{v}$). The parameters $k$, $r_0$, $p$, and $z$ show moderate influence. In contrast, $\epsilon_{e}$ and $\epsilon_{B}$ have negligible impact on the PD because polarization is geometric in origin and insensitive to the overall luminosity normalization. The parameters $\Delta_0$, $r_{b}$, and $\eta$ are also of limited importance: $\Delta_0$ and $r_{b}$ do not significantly affect the afterglow dynamics, while $\eta$ mainly influences the very early afterglow and becomes unimportant at the times of interest.

Degeneracy and the treatment of $\epsilon_{e}$ and $\epsilon_{B}$:
A natural concern is whether fixing $\epsilon_{e}$ and $\epsilon_{B}$ would bias the Monte Carlo results of Section~\ref{sec:4}. In synchrotron afterglow models degeneracy exists among $E_{iso}$, the ambient density, $\epsilon_{e}$, and $\epsilon_{B}$ in determining the flux normalization \citep{2013ApJ...776..120Y, 2024MNRAS.527.6752G}. Because of this degeneracy, the observational data alone provide only very weak constraints on $\epsilon_{e}$ and $\epsilon_{B}$ individually, and their fitted values would have large uncertainties and low credibility. Fixing them is therefore preferable to treating them as free parameters whose poorly constrained distributions could introduce spurious scatter into the Monte Carlo results. The K--S test performed in Section~\ref{sec:4} confirms that, with $\epsilon_{e}$ and $\epsilon_{B}$ fixed to their fiducial values, the simulated flux distributions remain consistent with the observed ones. Together with the fact that $\epsilon_{e}$ and $\epsilon_{B}$ have negligible impact on the PD, fixing them is both statistically well-motivated and physically justified.

\begin{figure*}[ht!]
\centering
\includegraphics[width=0.48\linewidth]{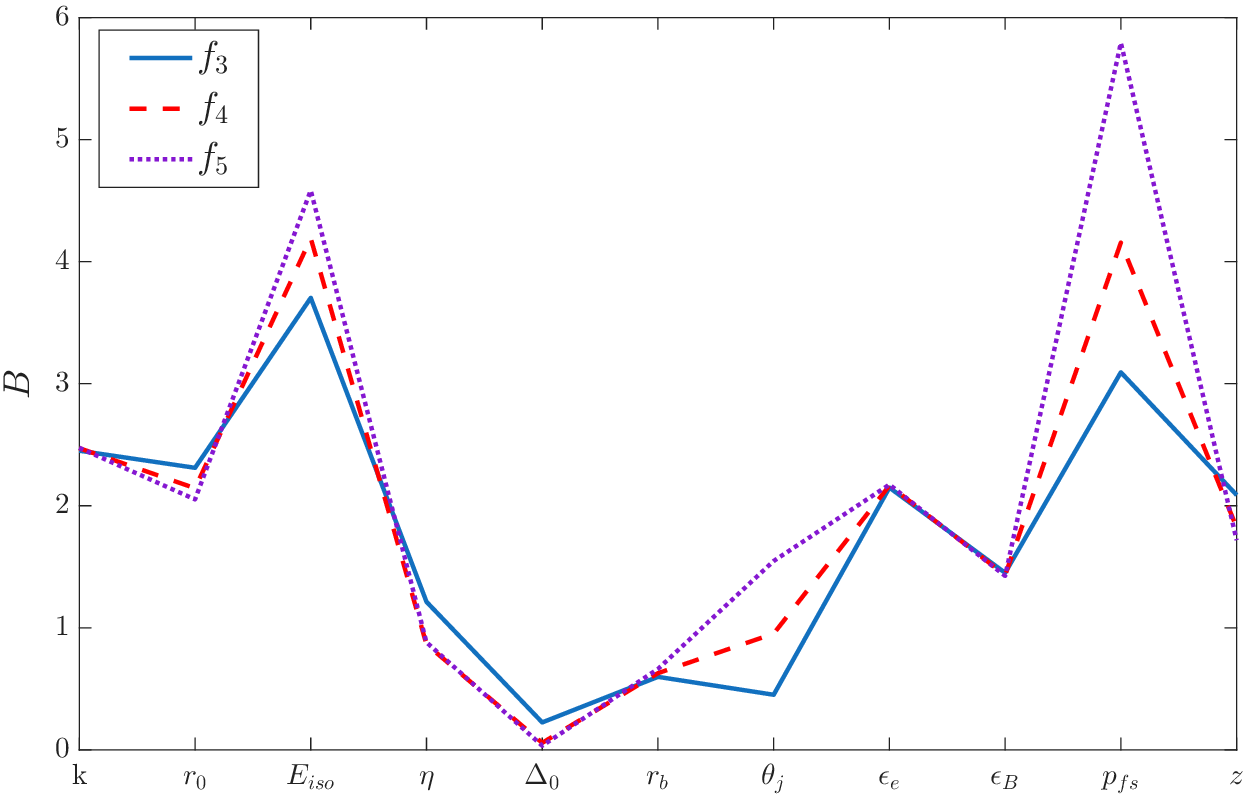}
\includegraphics[width=0.48\linewidth]{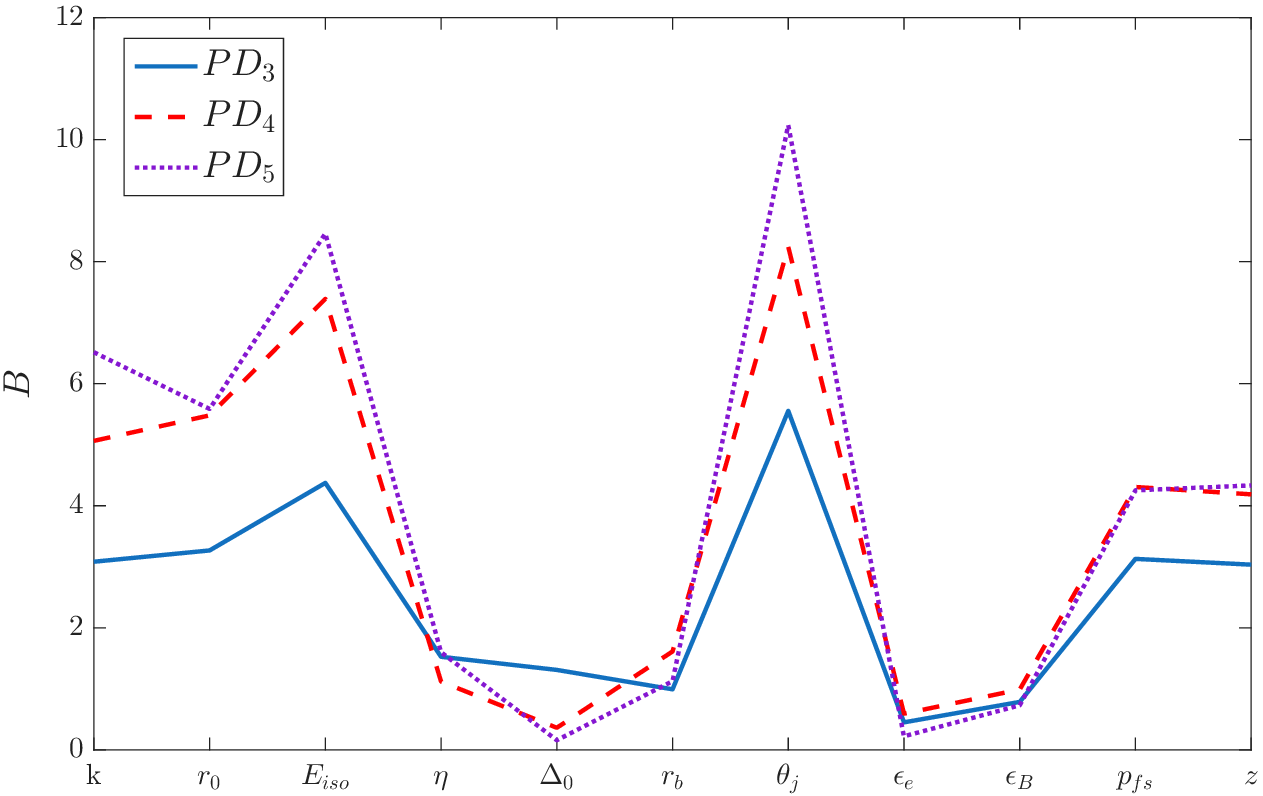}
\caption{Composite sensitivity index $B_i = \sqrt{\mu_i^{*2} + \sigma_i^2}$ for eleven model parameters. A larger $B_i$ indicates a greater overall influence of the corresponding  parameter on the predicted results of the model.
\textit{Left:} flux density at 10~keV. \textit{Right:} PD.
Blue solid, red dashed, and purple dotted lines correspond to $t_{obs} = 10^3$~s ($f_3$, ${PD}_3$), $10^4$~s ($f_4$, ${PD}_4$), and $10^5$~s ($f_5$, ${PD}_5$), respectively.}
\label{fig:2}
\end{figure*}

Final parameter selection:
Based on the screening, we retain $\{E_{iso}, \theta_{j}, k, r_0, p, z, \eta, q\}$ as the key parameters for the Monte Carlo simulations of Section~\ref{sec:4}, while $\{\Delta_0, r_{b}, \epsilon_{e}, \epsilon_{B}\}$ are fixed to their fiducial values. Fig.~\ref{fig:2} shows the composite index $B_i = \sqrt{\mu_i^{*2} + \sigma_i^2}$ for eleven parameters, evaluated separately for the flux density (left) and the PD (right) at the three representative observer times. The ranking confirms the parameter selection described above. We note that $B_i$ serves only as a convenient scalar metric for ranking and is not part of the original Morris framework \citep{Morris01051991, CAMPOLONGO20071509}.

\section{Detection prospects of the standard GRB X-ray afterglow by eXTP/PFA}
\label{sec:4}

Targeting neutron stars, magnetars, and black holes, the eXTP provides simultaneous spectral and timing data (0.5--10 keV) alongside with the polarimetric data (2--8 keV), enabling groundbreaking studies \citep{2025SCPMA..6819502Z}.
\par The PFA consists of a set of 3 X-ray telescopes with an angular resolution better than $30^{\prime\prime}$ (HPD) in a $9.8^{\prime} \times 9.8^{\prime}$ FoV, with a total effective area of $250~\text{cm}^2$ at 3 keV. It will be equipped with imaging gas pixel photoelectric polarimeters. PFA will be sensitive to energetic photons in the energy range 2--8 keV. For additional technical details, refer to \cite{2025SCPMA..6819502Z}. This information helps to reveal the jet structure, magnetic field configuration, and properties of the circumburst environment. For a typical afterglow flux ($10^{-9}~\text{erg}~\text{cm}^{-2}~\text{s}^{-1}$, 1000 s), PFA yields a Minimum Detectable Polarization (MDP) of $\sim 4.21\%$ at the 2--8 keV band \citep{2025SCPMA..6819506Y}.

\par We select a sample of GRB X-ray afterglows from bursts jointly detected by Fermi GBM and Swift XRT with measured spectroscopic redshifts and with X-ray data coverage near $10^3$~s and $10^4$~s. The X-ray light curves are taken from the Swift XRT repository \citep{2007A&A...469..379E}, and the prompt emission parameters from the Fermi GBM catalogs \citep{2014ApJS..211...12G, 2014ApJS..211...13V, 2016ApJS..223...28N, 2020ApJ...893...46V}. For each burst, we fit the 10~keV light curve with a power-law model $F_\nu(t) \propto t^{-\alpha}$ using weighted least squares in logarithmic space. Only bursts whose light curves are well described by this model are retained.
And the isotropic energy $E_{iso}$ was calculated using the Amati correlation \citep{2006MNRAS.372..233A}:
\begin{equation}
E_{p, i} = 95\times E_{iso}^{0.49},
\end{equation}
Here, $E_{p, i} = E_{peak} \times (1+z)$ denotes the peak energy in the burst source frame. 
Our final sample consists of 30 GRB afterglows, listed in Table~\ref{tab:sample}. For each burst, the 10~keV flux densities at $10^3$~s and $10^4$~s are obtained by evaluating the best-fit power law, denoted as $f_3$ and $f_4$, respectively.
The distributions of $f_3$ and $f_4$ serve as the observational benchmark for validating the Monte Carlo simulations.

\begin{table}
\caption{Sample of GRB X-ray afterglows.}
\label{tab:sample}
\centering
\setlength{\tabcolsep}{15pt}
\begin{tabular}{l c c}
\hline
GRB & $z$ & $\log_{10}(E_{iso}/{erg})$ \\
\hline
GRB\,080905B & 2.3740 & 53.6589 \\
GRB\,080916A & 0.6890 & 52.5617 \\
GRB\,081008   & 1.9685 & 53.4643 \\
GRB\,081221   & 2.2600 & 52.9684 \\
GRB\,081222   & 2.7700 & 53.5631 \\
GRB\,090102   & 1.5470 & 54.1565 \\
GRB\,090424   & 0.5440 & 52.8470 \\
GRB\,090519   & 3.8500 & 55.9078 \\
GRB\,091020   & 1.7100 & 53.6596 \\
GRB\,100413A  & 3.9000 & 55.4660 \\
GRB\,110818A  & 3.3600 & 53.8958 \\
GRB\,111107A  & 2.8930 & 53.9244 \\
GRB\,120712A  & 4.1745 & 52.5462 \\
GRB\,120922A  & 3.1000 & 52.3905 \\
GRB\,130420A  & 1.2970 & 52.2165 \\
GRB\,130612A  & 2.0060 & 51.9207 \\
GRB\,140206A  & 2.7300 & 53.3812 \\
GRB\,141220A  & 1.3195 & 53.3072 \\
GRB\,141221A  & 1.4520 & 52.7675 \\
GRB\,150301B  & 1.5169 & 53.3992 \\
GRB\,150403A  & 2.0600 & 54.3275 \\
GRB\,161017A  & 2.0127 & 53.7952 \\
GRB\,170705A  & 2.0100 & 53.0033 \\
GRB\,180205A  & 1.4090 & 51.8993 \\
GRB\,190114C  & 0.4200 & 54.5877 \\
GRB\,190324A  & 1.1715 & 52.9987 \\
GRB\,201021C  & 1.0700 & 52.7828 \\
GRB\,220101A  & 4.6100 & 54.3609 \\
GRB\,221226B  & 2.6940 & 52.9615 \\
GRB\,230506C  & 3.7000 & 53.5175 \\
\hline
\end{tabular}
\end{table}

\subsection{Statistical distributions}
\label{sec:distributions}

Based on the Morris sensitivity analysis, eight parameters are retained for Monte Carlo sampling:
$\{E_{iso}, \theta_{j}, k, r_0, p, z, \eta, q\}$. 
The remaining parameters are fixed:
$\Delta_0 = 1 \times 10^{10}~\rm cm$,
$r_{b} = 10^{13}~\rm cm$,
$\epsilon_{e} = 0.1$,
$\epsilon_{B} = 0.01$.

\noindent{$k$}: follows a normal distribution $\mathcal{N}(0.96,\;0.31^2)$ within a range of $[0, 3)$. The distribution is obtained by applying maximum-likelihood estimation to the set of $k$ values derived from the fitting of the 19-GRB afterglow  \citet{2013ApJ...776..120Y}.

\noindent{$r_0$}: follows a log-uniform prior distribution over $[10^{15}, 10^{18}]~\rm cm$. Owing to the degeneracy $n_0 r_0^k = {\rm const}$, $r_0$ cannot be independently constrained by afterglow light curves, we therefore adopt this conservative prior range.

\noindent{$p$}: its 68\% and 95\% credible intervals and the median are taken from the MCMC posterior reported by \citet{2015ApJ...799....3R}: its range is $[2.08, 2.78]$ for 68\% credible interval and is $[2.03, 4.05]$ for 95\% with a median value of $2.30$. We sample the random numbers of $p$ via quantile-interpolated inverse-CDF within 95\% credible interval.

\noindent{$z$}: follows a gamma distribution with shape $a = 3.61$ and scale $b = 0.62$, truncated at $z_{\rm max} = 6$. The distribution of $z$ is obtained via maximum-likelihood estimation from our selected Fermi--Swift sample listed in Table \ref{tab:sample}.

\noindent{$q$}: its 68\% and 95\% credible intervals and the median are taken from the MCMC posterior reported by \citet{2015ApJ...799....3R}: its range is $[0.26, 0.73]$ for 68\% credible interval and is $[0.16, 0.75]$ for 95\% with a median value of $0.57$. We sample the random number of $q$ via quantile-interpolated inverse-CDF within the 95\% credible interval.

\noindent{$E_{iso}$, $\theta_{j}$, and $\eta$}: 
$E_{iso}$ follows a log-normal distribution, $\log_{10}(E_{iso}/{erg}) \sim \mathcal{N}(53.41,\; 0.93^2)$, which is obtained via maximum-likelihood estimation from our Fermi--Swift sample listed in Table \ref{tab:sample}.
$\theta_{j}$ is drawn from the MCMC posterior reported by \citet{2015ApJ...799....3R}: 68\% credible interval is $[0.056, 0.33]~\rm rad$, 95\% credible interval is $[0.055, 0.42]~\rm rad$, and the median value is $0.097~\rm rad$. We sample its random number via quantile-interpolated inverse-CDF within the 95\% credible interval.
These two parameters ($E_{iso}$ and $\theta_{j}$) are sampled jointly via a Gaussian copula with correlation $\rho_c \approx -0.98$, calibrated to reproduce the \citet{2004ApJ...616..331G} relation: the geometry-corrected energy $E_\gamma \approx E_{iso}\theta_{j}^2/2$ is tightly clustered with an intrinsic scatter of $\sigma(\log E_\gamma) = 0.6~\rm dex$ ($\mathrm{Var}[\log E_{iso} + 2\log_{10}\theta_{j}] \approx 0.36~\rm dex^2$).
Finally, $\eta$ is conditionally sampled from $E_{iso}$ using the \citet{2010ApJ...725.2209L} relation, $\log_{10}\eta = 2.26 + 0.25\log_{10}(E_{iso}/10^{52}\,{erg}) + \varepsilon$, with intrinsic scatter $\varepsilon \sim \mathcal{N}(0, 0.11^2)$ ($\sigma(\log\eta) \approx 0.11~\rm dex$).

\subsection{Monte Carlo simulations}
We generate $10^3$ GRB afterglows by randomly sampling the parameter distributions specified in Section~\ref{sec:distributions}. For each event, the afterglow model of Section~2 is evaluated to obtain the flux density and PD at three representative observer times: $t_{obs} = 10^3~\rm s$, $10^4~\rm s$, and $10^5~\rm s$.

\begin{figure}[t]
\centering
\includegraphics[width=\columnwidth]{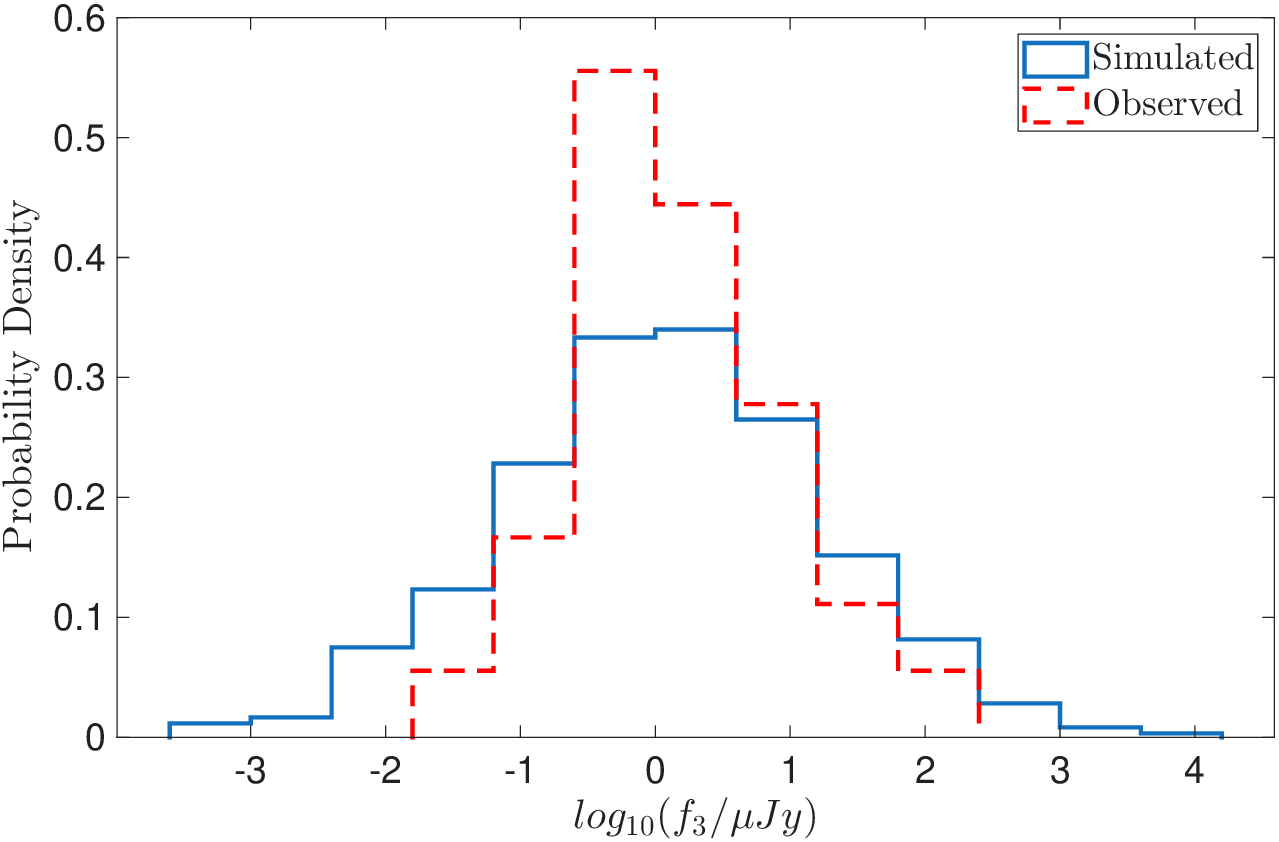}
\includegraphics[width=\columnwidth]{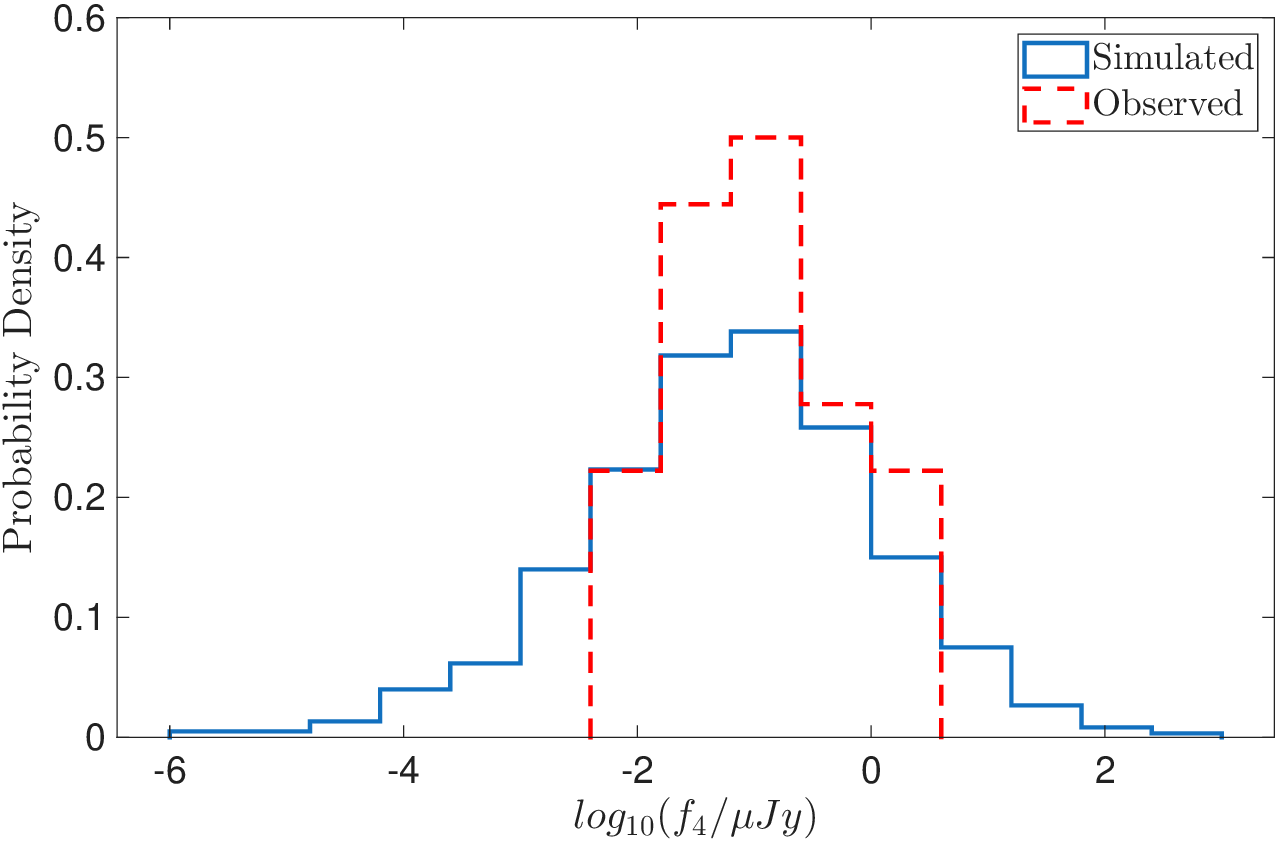}
\caption{
Probability density distributions of the flux densities of standard GRB afterglow at 10~keV at
$t_{obs}=10^{3}~\rm s$ (upper panel; K--S $p = 0.29$) and
$t_{obs}=10^{4}~\rm s$ (lower panel; K--S $p = 0.18$).
The blue solid stairs represent the simulated distribution and
the red dashed stairs are the observed distribution from the
Fermi--Swift sample \citep{2007A&A...469..379E, 2014ApJS..211...12G, 2014ApJS..211...13V, 2016ApJS..223...28N, 2020ApJ...893...46V}.
}
\label{fig:ks}
\end{figure}

\noindent{Model validation at 10~keV:} 
Before assessing polarization detectability, we validate the simulation pipeline by comparing the simulated flux density distributions at 10~keV with the observed ones from the Fermi--Swift sample. Fig.~\ref{fig:ks} presents the probability density distributions of the simulated and observed fluxes at $t_{obs}=10^3~\rm s$ and $10^4~\rm s$. A two-sample Kolmogorov--Smirnov (K--S) test yields $p = 0.29$ at $t_{obs}=10^3~\rm s$ and $p = 0.18$ at $t_{\rm obs}=10^4~\rm s$. Both $p$-values exceed $0.05$, indicating that the null hypothesis---that the simulated and observed flux distributions are drawn from the same parent population---cannot be rejected at the 95\% confidence level. This consistency validates the parameter distributions and the overall simulation methodology.

\begin{figure}[t]
\centering
\includegraphics[width=\columnwidth]{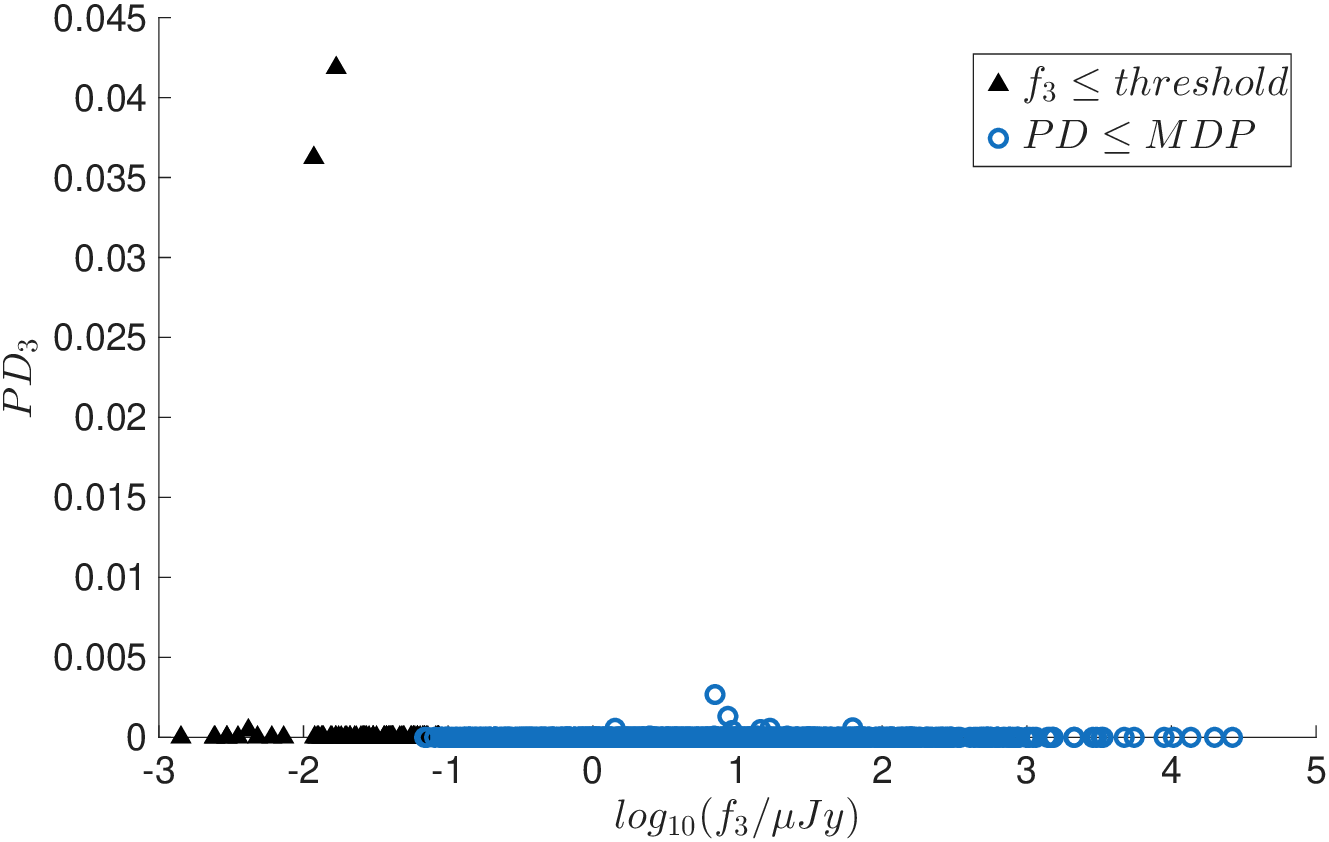}
\includegraphics[width=\columnwidth]{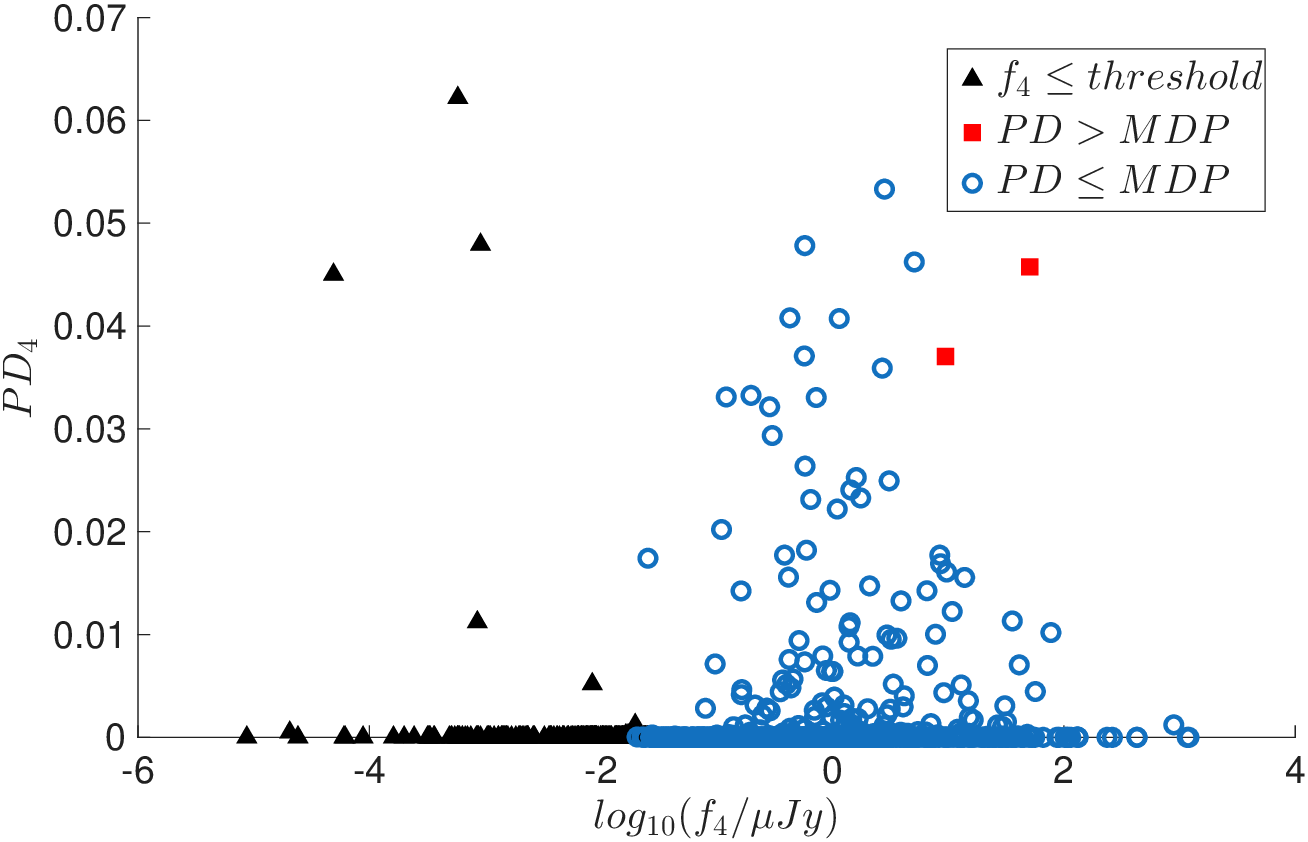}
\includegraphics[width=\columnwidth]{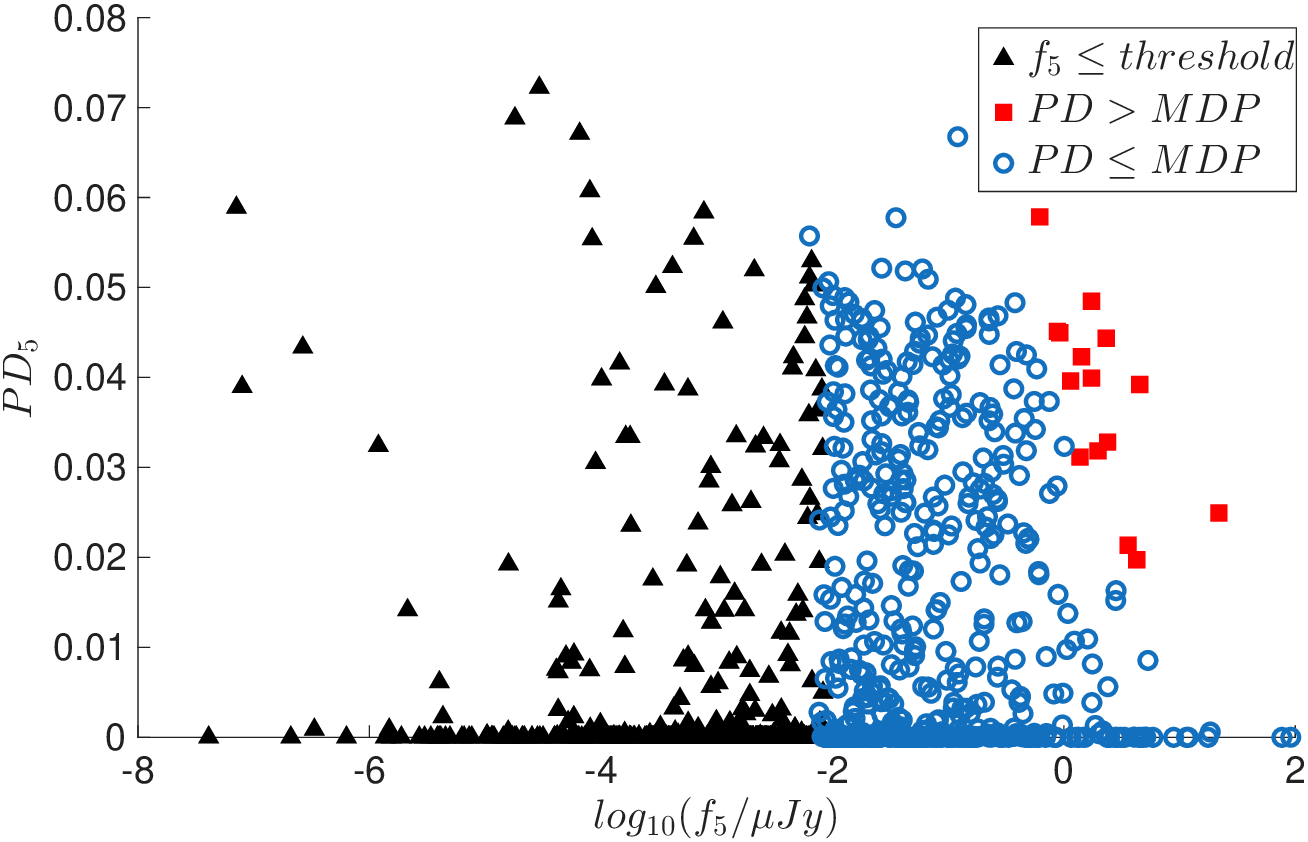}
\caption{
PD versus 2--8~keV flux density for the $10^3$
simulated events at $t_{obs}=10^{3}~\rm s$ (upper panel),
$t_{obs}=10^{4}~\rm s$ (middle panel), and
$t_{obs}=10^{5}~\rm s$ (lower panel).
Black triangles mark events whose flux falls below the eXTP/PFA
sensitivity threshold.
For events above the flux threshold, blue circles indicate
$PD \le MDP$ and red squares indicate
$PD > MDP$.
The numbers of detectable events (red squares) at the three epochs
are 0, 2, and 15, respectively.
}
\label{fig:MDP}
\end{figure}

\noindent{Polarization detectability at 2--8~keV:}
We now assess the detectability of the simulated events by eXTP/PFA. For each event, the 2--8~keV integrated flux density and the corresponding PD are computed at the same three observer times. Two criteria must be satisfied for a detection: (i) the calculated 2--8~keV flux must exceed the eXTP/PFA sensitivity threshold; (ii) the calculated PD must exceed the MDP at the 99\% confidence level \citep{2025SCPMA..6819502Z, 2025SCPMA..6819505G, 2025SCPMA..6819506Y}. For the MDP calculation, the exposure covers the interval $[t_{obs},\, 2t_{obs}]$, i.e., $[10^3,\, 2\times 10^3]~\rm s$, $[10^4,\, 2\times 10^4]~\rm s$, and $[10^5,\, 2\times 10^5]~\rm s$, respectively.
Fig.~\ref{fig:MDP} displays the PD versus 2--8~keV flux density for all simulated events. At $t_{obs} = 10^3~\rm s$, no event satisfies both criteria simultaneously. At $t_{obs} = 10^4~\rm s$, 2 events (0.2\%) lie above the MDP. At $t_{obs} = 10^5~\rm s$, 15 events (1.5\%) are detectable.
The apparent increase in the number of detectable events with time may seem counterintuitive given the monotonic flux decay of the afterglow. This trend arises because, for many simulated events, the PD peaks at late times, whereas at early times the polarization signal remains well below the MDP despite the higher flux. The overall event rate remains low ($\lesssim 1.5\%$ across all epochs), consistent with the intrinsically weak polarization produced by a random magnetic field confined to the shock plane.

\section{The exceptional case of GRB 221009A}
\label{sec:5}
\par GRB 221009A, an extraordinarily bright GRB detected on October 9, 2022 by missions including the Fermi Gamma-ray Space Telescope, the Swift Observatory and GECAM, became a key target for polarization studies due to its unprecedented luminosity \citep{2023ApJ...952L..42L, 2023ApJ...946L..24W, 2024SCPMA..6789511Z}.
Consequently, the Imaging X-ray Polarimetry Explorer (IXPE) performed the first-ever measurement of its 2--8 keV polarized afterglow \citep{2023ApJ...946L..21N}.
\par
\begin{figure}[ht!]
\centering
\includegraphics[width=\hsize]{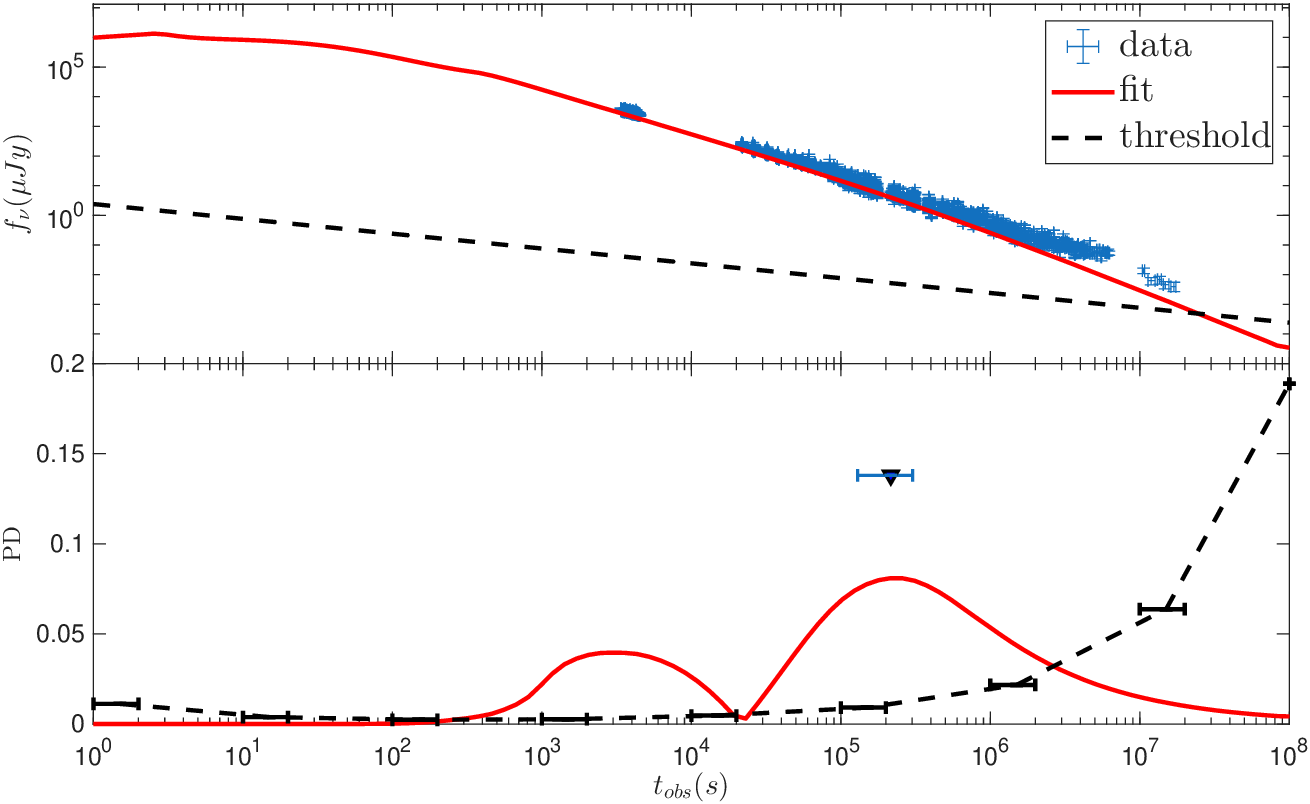}
\caption{Light curve (upper panel) and PD curve (lower panel) for fitting GRB 221009A afterglow observations at 2--8 keV, the blue points, red-solid line and black-dashed line represent the observational data, the best-fit curve, and the threshold and inverted triangles represent the upper limits of the PD values respectively. For comparison with the MDP of eXTP/PFA, the absolute value of the PD is shown.
The flux density data at 10 keV are taken from Swift \citep{2007A&A...469..379E} and the polarization data at 2--8 keV are taken from \citet{2023ApJ...946L..21N}.
\label{fig:7}}
\end{figure}

 In this section, we apply the model from \cite{2023ApJ...952...31L} to fit the afterglow observations of GRB 221009A, with the fitting results presented in Fig.~\ref{fig:7}.

The best-fit parameters are as follows: $E_{iso} = 1 \times 10^{55}~\text{erg}$, $\eta = 700$, $\epsilon_{e,fs} = 0.1$, $\epsilon_{B,fs} = 0.01$, $p_{fs} = 2.46$, $r_0 = 10^{17}~\text{cm}$, $n_0 = 9~\text{cm}^{-3}$, $k = 1.65$, $z = 0.151$ \citep{2022GCN.32648....1D}, $\theta_j = 0.026~\text{rad}$ \citep{2023ApJ...946L..21N}, $q = 2/3$ \citep{1999MNRAS.309L...7G} and the contribution from the reverse shock is neglected.
The results clearly show that throughout the observational period from $10^3$ s to $10^6$ s, the afterglow PD of GRB 221009A remains almost entirely above the MDP of eXTP/PFA. This indicates that such exceptionally bright GRBs can produce detectable PD signals over these extended durations.

\section{Conclusions and discussion}
\label{sec:6}

In this paper, we have systematically investigated the polarization properties of GRB X-ray afterglows in the 2--8~keV band and assessed their detectability with eXTP/PFA.
A Morris global sensitivity analysis was performed to identify the parameters that dominate the predicted polarization, retaining eight parameters while fixing the remaining four to fiducial values.
Statistical distributions for the retained parameters were constructed from a carefully selected Fermi--Swift sample and from the literature; in particular, $E_{iso}$, $\theta_{j}$, and $\eta$ were sampled jointly via a Gaussian copula to reproduce the \citet{2004ApJ...616..331G} and \citet{2010ApJ...725.2209L} correlations.
Monte Carlo simulations of $10^3$ afterglows were then performed and validated against the observed 10~keV flux distributions (K--S $p = 0.29$ at $10^3~\rm s$ and $p = 0.18$ at $10^4~\rm s$), confirming that the parameter choices are consistent with the data.
\par The Monte Carlo results indicate a low polarization event rate for standard GRB X-ray afterglows with eXTP/PFA.
At $t_{obs} = 10^3~\rm s$, none of the $10^3$ simulated afterglows simultaneously satisfies the flux-threshold and MDP criteria.
At $t_{obs} = 10^4~\rm s$, 2 events (0.2\%) become detectable, and at $t_{obs} = 10^5~\rm s$, 15 events (1.5\%) exceed the MDP.
The modest increase at late times, despite the overall flux decay, arises because the PD naturally peaks near the jet break for many lines of sight, partially compensating for the rising MDP.
The overall event rate of $\lesssim 1.5\%$ reflects the intrinsically low polarization produced by a random magnetic field confined to the shock plane.
\par For exceptionally bright events such as GRB~221009A, the outlook is markedly different.
Our model fit shows that its PD remains above the eXTP/PFA MDP over the full interval $10^3$--$10^6~\rm s$, demonstrating that eXTP/PFA can capture nearly the entire polarization evolution for such rare, luminous bursts.

\par We emphasize that the predictions presented here apply specifically to the standard external shock afterglow segments.
The X-ray band afterglow frequently exhibits additional components---flares and shallow-decay plateaus, which violates the standard closure relations---that depart from the standard model \citep{2006ApJ...642..354Z, 2006ApJ...642..389N, 2007ApJ...662.1093W}, and whose polarization signatures have been investigated in a growing body of work \citep{2016ApJ...816...73L, 2018ApJ...862..115G, 2025ApJ...989..172W, 2026ApJ...996...33W}.
Whether these components yield a detectable polarization depends sensitively on the underlying emission model: flares with large-scale ordered magnetic fields may produce high PD, whereas those with random magnetic fields predict near-zero polarization \citep{2025ApJ...989..172W}. The relativistic wind bubble model for plateaus predicts PD well above the eXTP/PFA threshold, while the structured ejecta model does not \citep{2026ApJ...996...33W}.
Our result shows that polarization detection of the standard GRB afterglow by eXTP/PFA is undeserving, except these extreme boat GRBs.

\begin{acknowledgements}  
We thank the anonemous referee for useful comments. This work is supported by the National Natural Science Foundation of China (grant No. 12473040, 12393812, 12041306, 12333007) and China's Space Origins Exploration Program.
\end{acknowledgements}
\bibliography{references}

@ARTICLE{1993ApJ...413L.101K,
       author = {{Kouveliotou}, Chryssa and {Meegan}, Charles A. and {Fishman}, Gerald J. and {Bhat}, Narayana P. and {Briggs}, Michael S. and {Koshut}, Thomas M. and {Paciesas}, William S. and {Pendleton}, Geoffrey N.},
        title = "{Identification of Two Classes of Gamma-Ray Bursts}",
      journal = {\apjl},
         year = 1993,
        month = aug,
       volume = {413},
        pages = {L101},
          doi = {10.1086/186969},
       adsurl = {https://ui.adsabs.harvard.edu/abs/1993ApJ...413L.101K}
}

@ARTICLE{2024ApJ...976...62Z,
       author = {{Zhu}, Si-Yuan and {Tam}, Pak-Hin Thomas},
        title = "{Unveiling the Progenitors of a Population of Likely Peculiar Gamma-Ray Bursts}",
      journal = {\apj},
         year = 2024,
        month = nov,
       volume = {976},
       number = {1},
          eid = {62},
        pages = {62},
          doi = {10.3847/1538-4357/ad813a},
archivePrefix = {arXiv},
       eprint = {2409.19719},
 primaryClass = {astro-ph.HE},
       adsurl = {https://ui.adsabs.harvard.edu/abs/2024ApJ...976...62Z}
}

@ARTICLE{1986ApJ...308L..43P,
       author = {{Paczynski}, B.},
        title = "{Gamma-ray bursters at cosmological distances}",
      journal = {\apjl},
         year = 1986,
        month = sep,
       volume = {308},
        pages = {L43-L46},
          doi = {10.1086/184740},
       adsurl = {https://ui.adsabs.harvard.edu/abs/1986ApJ...308L..43P}
}

@ARTICLE{1993ApJ...405..273W,
       author = {{Woosley}, S.~E.},
        title = "{Gamma-Ray Bursts from Stellar Mass Accretion Disks around Black Holes}",
      journal = {\apj},
         year = 1993,
        month = mar,
       volume = {405},
        pages = {273},
          doi = {10.1086/172359},
       adsurl = {https://ui.adsabs.harvard.edu/abs/1993ApJ...405..273W}
}

@ARTICLE{2007ApJ...655L..25Z,
       author = {{Zhang}, Bing and {Zhang}, Bin-Bin and {Liang}, En-Wei and {Gehrels}, Neil and {Burrows}, David N. and {M{\'e}sz{\'a}ros}, Peter},
        title = "{Making a Short Gamma-Ray Burst from a Long One: Implications for the Nature of GRB 060614}",
      journal = {\apjl},
         year = 2007,
        month = jan,
       volume = {655},
       number = {1},
        pages = {L25-L28},
          doi = {10.1086/511781},
archivePrefix = {arXiv},
       eprint = {astro-ph/0612238},
 primaryClass = {astro-ph},
       adsurl = {https://ui.adsabs.harvard.edu/abs/2007ApJ...655L..25Z}
}

@ARTICLE{2003ApJ...599L..95M,
       author = {{Mazzali}, Paolo A. and {Deng}, Jinsong and {Tominaga}, Nozomu and {Maeda}, Keiichi and {Nomoto}, Ken'ichi and {Matheson}, Thomas and {Kawabata}, Koji S. and {Stanek}, Krzysztof Z. and {Garnavich}, Peter M.},
        title = "{The Type Ic Hypernova SN 2003dh/GRB 030329}",
      journal = {\apjl},
         year = 2003,
        month = dec,
       volume = {599},
       number = {2},
        pages = {L95-L98},
          doi = {10.1086/381259},
archivePrefix = {arXiv},
       eprint = {astro-ph/0309555},
 primaryClass = {astro-ph},
       adsurl = {https://ui.adsabs.harvard.edu/abs/2003ApJ...599L..95M}
}

@ARTICLE{2017ApJ...848L..14G,
       author = {{Goldstein}, A. and {Veres}, P. and {Burns}, E. and {Briggs}, M.~S. and {Hamburg}, R. and {Kocevski}, D. and {Wilson-Hodge}, C.~A. and {Preece}, R.~D. and {Poolakkil}, S. and {Roberts}, O.~J. and et al.},
        title = "{An Ordinary Short Gamma-Ray Burst with Extraordinary Implications: Fermi-GBM Detection of GRB 170817A}",
      journal = {\apjl},
         year = 2017,
        month = oct,
       volume = {848},
       number = {2},
          eid = {L14},
        pages = {L14},
          doi = {10.3847/2041-8213/aa8f41},
archivePrefix = {arXiv},
       eprint = {1710.05446},
 primaryClass = {astro-ph.HE},
       adsurl = {https://ui.adsabs.harvard.edu/abs/2017ApJ...848L..14G}
}

@ARTICLE{2017ApJ...848L..13A,
       author = {{Abbott}, B.~P. and {Abbott}, R. and {Abbott}, T.~D. and {Acernese}, F. and {Ackley}, K. and {Adams}, C. and {Adams}, T. and {Addesso}, P. and {Adhikari}, R.~X. and {Adya}, V.~B. and et al.},
        title = "{Gravitational Waves and Gamma-Rays from a Binary Neutron Star Merger: GW170817 and GRB 170817A}",
      journal = {\apjl},
         year = 2017,
        month = oct,
       volume = {848},
       number = {2},
          eid = {L13},
        pages = {L13},
          doi = {10.3847/2041-8213/aa920c},
archivePrefix = {arXiv},
       eprint = {1710.05834},
 primaryClass = {astro-ph.HE},
       adsurl = {https://ui.adsabs.harvard.edu/abs/2017ApJ...848L..13A}
}

@ARTICLE{2018PhRvL.120x1103L,
       author = {{Lazzati}, Davide and {Perna}, Rosalba and {Morsony}, Brian J. and {Lopez-Camara}, Diego and {Cantiello}, Matteo and {Ciolfi}, Riccardo and {Giacomazzo}, Bruno and {Workman}, Jared C.},
        title = "{Late Time Afterglow Observations Reveal a Collimated Relativistic Jet in the Ejecta of the Binary Neutron Star Merger GW170817}",
      journal = {\prl},
         year = 2018,
        month = jun,
       volume = {120},
       number = {24},
          eid = {241103},
        pages = {241103},
          doi = {10.1103/PhysRevLett.120.241103},
archivePrefix = {arXiv},
       eprint = {1712.03237},
 primaryClass = {astro-ph.HE},
       adsurl = {https://ui.adsabs.harvard.edu/abs/2018PhRvL.120x1103L}
}

@ARTICLE{2015NatPh..11..173H,
       author = {{Huntington}, C.~M. and {Fiuza}, F. and {Ross}, J.~S. and {Zylstra}, A.~B. and {Drake}, R.~P. and {Froula}, D.~H. and {Gregori}, G. and {Kugland}, N.~L. and {Kuranz}, C.~C. and {Levy}, M.~C. and et al.},
        title = "{Observation of magnetic field generation via the Weibel instability in interpenetrating plasma flows}",
      journal = {Nature Physics},
         year = 2015,
        month = feb,
       volume = {11},
       number = {2},
        pages = {173-176},
          doi = {10.1038/nphys3178},
archivePrefix = {arXiv},
       eprint = {1310.3337},
 primaryClass = {astro-ph.HE},
       adsurl = {https://ui.adsabs.harvard.edu/abs/2015NatPh..11..173H}
}

@ARTICLE{1977MNRAS.179..433B,
       author = {{Blandford}, R.~D. and {Znajek}, R.~L.},
        title = "{Electromagnetic extraction of energy from Kerr black holes.}",
      journal = {\mnras},
         year = 1977,
        month = may,
       volume = {179},
        pages = {433-456},
          doi = {10.1093/mnras/179.3.433},
       adsurl = {https://ui.adsabs.harvard.edu/abs/1977MNRAS.179..433B}
}

@ARTICLE{2001A&A...369..694S,
       author = {{Spruit}, H.~C. and {Daigne}, F. and {Drenkhahn}, G.},
        title = "{Large scale magnetic fields and their dissipation in GRB fireballs\}",
      journal = {\aap},
     keywords = {GAMMA-RAY BURSTS, MAGNETIC FIELDS, RADIATION MECHANISMS: NONTHERMAL, SHOCK WAVES, Astrophysics},
         year = 2001,
        month = apr,
       volume = {369},
        pages = {694-705},
          doi = {10.1051/0004-6361:20010131},
archivePrefix = {arXiv},
       eprint = {astro-ph/0004274},
 primaryClass = {astro-ph},
       adsurl = {https://ui.adsabs.harvard.edu/abs/2001A&A...369..694S},
      adsnote = {Provided by the SAO/NASA Astrophysics Data System}
}

@ARTICLE{2013ApJ...776..120Y,
       author = {{Yi}, Shuang-Xi and {Wu}, Xue-Feng and {Dai}, Zi-Gao},
        title = "{Early Afterglows of Gamma-Ray Bursts in a Stratified Medium with a Power-law Density Distribution}",
      journal = {\apj},
         year = 2013,
        month = oct,
       volume = {776},
       number = {2},
          eid = {120},
        pages = {120},
          doi = {10.1088/0004-637X/776/2/120},
archivePrefix = {arXiv},
       eprint = {1308.6095},
 primaryClass = {astro-ph.HE},
       adsurl = {https://ui.adsabs.harvard.edu/abs/2013ApJ...776..120Y}
}

@ARTICLE{1998ApJ...497L..17S,
       author = {{Sari}, Re'em and {Piran}, Tsvi and {Narayan}, Ramesh},
        title = "{Spectra and Light Curves of Gamma-Ray Burst Afterglows}",
      journal = {\apjl},
         year = 1998,
        month = apr,
       volume = {497},
       number = {1},
        pages = {L17-L20},
          doi = {10.1086/311269},
archivePrefix = {arXiv},
       eprint = {astro-ph/9712005},
 primaryClass = {astro-ph},
       adsurl = {https://ui.adsabs.harvard.edu/abs/1998ApJ...497L..17S}
}

@ARTICLE{2000ApJ...543...90H,
       author = {{Huang}, Y.~F. and {Gou}, L.~J. and {Dai}, Z.~G. and {Lu}, T.},
        title = "{Overall Evolution of Jetted Gamma-Ray Burst Ejecta}",
      journal = {\apj},
         year = 2000,
        month = nov,
       volume = {543},
       number = {1},
        pages = {90-96},
          doi = {10.1086/317076},
archivePrefix = {arXiv},
       eprint = {astro-ph/9910493},
 primaryClass = {astro-ph},
       adsurl = {https://ui.adsabs.harvard.edu/abs/2000ApJ...543...90H}
}

@ARTICLE{2004MNRAS.354...86R,
       author = {{Rossi}, Elena M. and {Lazzati}, Davide and {Salmonson}, Jay D. and {Ghisellini}, Gabriele},
        title = "{The polarization of afterglow emission reveals {\ensuremath{\gamma}}-ray bursts jet structure}",
      journal = {\mnras},
         year = 2004,
        month = oct,
       volume = {354},
       number = {1},
        pages = {86-100},
          doi = {10.1111/j.1365-2966.2004.08165.x},
archivePrefix = {arXiv},
       eprint = {astro-ph/0401124},
 primaryClass = {astro-ph},
       adsurl = {https://ui.adsabs.harvard.edu/abs/2004MNRAS.354...86R}
}

@ARTICLE{2023ApJ...952...31L,
       author = {{Lan}, Mi-Xiang and {Wu}, Xue-Feng and {Dai}, Zi-Gao},
        title = "{Afterglow Polarizations in a Stratified Medium with Effect of the Equal Arrival Time Surface}",
      journal = {\apj},
         year = 2023,
        month = jul,
       volume = {952},
       number = {1},
          eid = {31},
        pages = {31},
          doi = {10.3847/1538-4357/acd6ef},
archivePrefix = {arXiv},
       eprint = {2305.10590},
 primaryClass = {astro-ph.HE},
       adsurl = {https://ui.adsabs.harvard.edu/abs/2023ApJ...952...31L}
}

@ARTICLE{2016ApJ...816...73L,
       author = {{Lan}, Mi-Xiang and {Wu}, Xue-Feng and {Dai}, Zi-Gao},
        title = "{Polarization Evolution of Early Optical Afterglows of Gamma-Ray Bursts}",
      journal = {\apj},
         year = 2016,
        month = jan,
       volume = {816},
       number = {2},
          eid = {73},
        pages = {73},
          doi = {10.3847/0004-637X/816/2/73},
archivePrefix = {arXiv},
       eprint = {1511.01582},
 primaryClass = {astro-ph.HE},
       adsurl = {https://ui.adsabs.harvard.edu/abs/2016ApJ...816...73L}
}

@ARTICLE{1999ApJ...520..641S,
       author = {{Sari}, Re'em and {Piran}, Tsvi},
        title = "{Predictions for the Very Early Afterglow and the Optical Flash}",
      journal = {\apj},
         year = 1999,
        month = aug,
       volume = {520},
       number = {2},
        pages = {641-649},
          doi = {10.1086/307508},
archivePrefix = {arXiv},
       eprint = {astro-ph/9901338},
 primaryClass = {astro-ph},
       adsurl = {https://ui.adsabs.harvard.edu/abs/1999ApJ...520..641S}
}

@ARTICLE{2003Natur.423..388W,
       author = {{Waxman}, Eli},
        title = "{Astronomy: New direction for {\ensuremath{\gamma}}-rays}",
      journal = {\nat},
         year = 2003,
        month = may,
       volume = {423},
       number = {6938},
        pages = {388-389},
          doi = {10.1038/423388a},
archivePrefix = {arXiv},
       eprint = {astro-ph/0305414},
 primaryClass = {astro-ph},
       adsurl = {https://ui.adsabs.harvard.edu/abs/2003Natur.423..388W}
}

@ARTICLE{1999MNRAS.309L...7G,
       author = {{Ghisellini}, Gabriele and {Lazzati}, Davide},
        title = "{Polarization light curves and position angle variation of beamed gamma-ray bursts}",
      journal = {\mnras},
         year = 1999,
        month = oct,
       volume = {309},
       number = {1},
        pages = {L7-L11},
          doi = {10.1046/j.1365-8711.1999.03025.x},
archivePrefix = {arXiv},
       eprint = {astro-ph/9906471},
 primaryClass = {astro-ph},
       adsurl = {https://ui.adsabs.harvard.edu/abs/1999MNRAS.309L...7G}
}

@ARTICLE{2024NatAs...8..134A,
       author = {{Arimoto}, Makoto and {Asano}, Katsuaki and {Kawabata}, Koji S. and {Toma}, Kenji and {Gill}, Ramandeep and {Granot}, Jonathan and {Ohno}, Masanori and {Takahashi}, Shuta and {Ogino}, Naoki and {Goto}, Hatsune and et al.},
        title = "{Gamma rays from a reverse shock with turbulent magnetic fields in GRB 180720B}",
      journal = {Nature Astronomy},
         year = 2024,
        month = jan,
       volume = {8},
       number = {1},
        pages = {134-144},
          doi = {10.1038/s41550-023-02119-1},
archivePrefix = {arXiv},
       eprint = {2310.04144},
 primaryClass = {astro-ph.HE},
       adsurl = {https://ui.adsabs.harvard.edu/abs/2024NatAs...8..134A}
}

@ARTICLE{2025SCPMA..6819502Z,
       author = {{Zhang}, Shuang-Nan and {Santangelo}, Andrea and {Xu}, Yupeng and {Feng}, Hua and {Lu}, Fangjun and {Chen}, Yong and {Ge}, Mingyu and {Nandra}, Kirpal and {Wu}, Xin and {Feroci}, Marco and {Hernanz}, Margarita and {Liu}, Congzhan and {He}, Huilin and {Wang}, Yusa and {Jiang}, Weichun and {Cui}, Weiwei and {Yang}, Yanji and {Wang}, Juan and {Li}, Wei and {Li}, Hong and {Du}, Yuanyuan and {Liu}, Xiaohua and {Meng}, Bin and {Wen}, Xiangyang and {Zhang}, Aimei and {Ma}, Jia and {Li}, Maoshun and {Li}, Gang and {Qi}, Liqiang and {Sun}, Jianchao and {Luo}, Tao and {Liu}, Hongwei and {Liu}, Xiaojing and {Zhang}, Fan and {Luo}, Laidan and {Zhu}, Yuxuan and {Zhao}, Zijian and {Sun}, Liang and {Yang}, Xiongtao and {Wu}, Qiong and {Jiang}, Jiechen and {Shi}, Haoli and {Liu}, Jiangtao and {Xu}, Yanbing and {Yang}, Sheng and {Zhang}, Laiyu and {Han}, Dawei and {Gao}, Na and {Huo}, Jia and {Zhang}, Ziliang and {Wang}, Hao and {Zhao}, Xiaofan and {Wang}, Shuo and {Li}, Zhenjie and {Bao}, Ziyu and {Liu}, Yaoguang and {Wang}, Ke and {Wang}, Na and {Wang}, Bo and {Wang}, Langping and {Wang}, Dianlong and {Ding}, Fei and {Sheng}, Lizhi and {Qiang}, Pengfei and {Yan}, Yongqing and {Liu}, Yongan and {Wu}, Zhenyu and {Liu}, Yichen and {Chen}, Hao and {Zhang}, Yacong and {Liu}, Hongbang and {Altmann}, Alexander and {Bechteler}, Thomas and {Burwitz}, Vadim and {Fiorini}, Carlo and {Friedrich}, Peter and {Meidinger}, Norbert and {Strecker}, Rafael and {Baldini}, Luca and {Bellazzini}, Ronaldo and {Bonino}, Raffaella and {Frass{\`a}}, Andrea and {Latronico}, Luca and {Maldera}, Simone and {Manfreda}, Alberto and {Minuti}, Massimo and {Pesce-Rollins}, Melissa and {Sgr{\`o}}, Carmelo and {Tugliani}, Stefano and {Pareschi}, Giovanni and {Basso}, Stefano and {Sironi}, Giorgia and {Spiga}, Daniele and {Tagliaferri}, Gianpiero and {Tykhonov}, Andrii and {Paltani}, St{\`e}phane and {Bozzo}, Enrico and {Tenzer}, Christoph and {Bayer}, J{\"o}rg and {Tuo}, Youli and {Liu}, Honghui and {Zhang}, Yonghe and {Cai}, Zhiming and {Liu}, Huaqiu and {Chen}, Wen and {Wang}, Chunhong and {He}, Tao and {Chen}, Yehai and {Qiu}, Chengbo and {Zhang}, Ye and {Feng}, Jianchao and {Zhu}, Xiaofei and {Zhou}, Heng and {Zheng}, Shijie and {Song}, Liming and {Wang}, Jinzhou and {Jia}, Shumei and {Jiang}, Zewen and {Li}, Xiaobo and {Zhao}, Haisheng and {Guan}, Ju and {Zhang}, Juan and {Li}, Chengkui and {Huang}, Yue and {Liao}, Jinyuan and {You}, Yuan and {Zhang}, Hongmei and {Wang}, Wenshuai and {Wang}, Shuang and {Ou}, Ge and {Hu}, Hao and {Shi}, Jingyan and {Cui}, Tao and {Jiang}, Xiaowei and {Cheng}, Yaodong and {Li}, Haibo and {Xu}, Yanjun and {Zane}, Silvia and {Bambi}, Cosimo and {Bu}, Qingcui and {Dall'Osso}, Simone and {Rosa}, Alessandra De and {Gou}, Lijun and {Guillot}, Sebastien and {Ji}, Long and {Li}, Ang and {Mao}, Jirong and {Patruno}, Alessandro and {Stratta}, Giulia and {Taverna}, Roberto and {Tsygankov}, Sergey and {Uttley}, Phil and {Watts}, Anna L. and {Wu}, Xuefeng and {Xu}, Renxin and {Yi}, Shuxu and {Zhang}, Guobao and {Zhang}, Liang and {Zhao}, Wen and {Zhou}, Ping},
        title = "{The enhanced X-ray Timing and Polarimetry mission{\textemdash}eXTP for launch in 2030}",
      journal = {Science China Physics, Mechanics, and Astronomy},
         year = 2025,
        month = sep,
       volume = {68},
       number = {11},
          eid = {119502},
        pages = {119502},
          doi = {10.1007/s11433-025-2786-6},
archivePrefix = {arXiv},
       eprint = {2506.08101},
 primaryClass = {astro-ph.HE},
       adsurl = {https://ui.adsabs.harvard.edu/abs/2025SCPMA..6819502Z}
}

@ARTICLE{2025SCPMA..6819506Y,
       author = {{Yi}, Shu-Xu and {Zhao}, Wen and {Xu}, Ren-Xin and {Wu}, Xue-Feng and {Stratta}, Giulia and {Dall'Osso}, Simone and {Xu}, Yan-Jun and {Santangelo}, Andrea and {Zane}, Silvia and {Zhang}, Shuang-Nan and et al.},
        title = "{Prospects for time-domain and multi-messenger science with eXTP}",
      journal = {Science China Physics, Mechanics, and Astronomy},
         year = 2025,
        month = sep,
       volume = {68},
       number = {11},
          eid = {119506},
        pages = {119506},
          doi = {10.1007/s11433-025-2782-2},
archivePrefix = {arXiv},
       eprint = {2506.08368},
 primaryClass = {astro-ph.HE},
       adsurl = {https://ui.adsabs.harvard.edu/abs/2025SCPMA..6819506Y}
}

@BOOK{1979rpa..book.....R,
       author = {{Rybicki}, George B. and {Lightman}, Alan P.},
        title = "{Radiative processes in astrophysics}",
         year = 1979,
       adsurl = {https://ui.adsabs.harvard.edu/abs/1979rpa..book.....R}
}

@ARTICLE{1999ApJ...524L..43S,
       author = {{Sari}, Re'em},
        title = "{Linear Polarization and Proper Motion in the Afterglow of Beamed Gamma-Ray Bursts}",
      journal = {\apjl},
         year = 1999,
        month = oct,
       volume = {524},
       number = {1},
        pages = {L43-L46},
          doi = {10.1086/312294},
archivePrefix = {arXiv},
       eprint = {astro-ph/9906503},
 primaryClass = {astro-ph},
       adsurl = {https://ui.adsabs.harvard.edu/abs/1999ApJ...524L..43S}
}

@ARTICLE{2007A&A...469..379E,
       author = {{Evans}, P.~A. and {Beardmore}, A.~P. and {Page}, K.~L. and {Tyler}, L.~G. and {Osborne}, J.~P. and {Goad}, M.~R. and {O'Brien}, P.~T. and {Vetere}, L. and {Racusin}, J. and {Morris}, D. and {Burrows}, D.~N. and {Capalbi}, M. and {Perri}, M. and {Gehrels}, N. and {Romano}, P.},
        title = "{An online repository of Swift/XRT light curves of {\ensuremath{\gamma}}-ray bursts}",
      journal = {\aap},
         year = 2007,
        month = jul,
       volume = {469},
       number = {1},
        pages = {379-385},
          doi = {10.1051/0004-6361:20077530},
archivePrefix = {arXiv},
       eprint = {0704.0128},
 primaryClass = {astro-ph},
       adsurl = {https://ui.adsabs.harvard.edu/abs/2007A&A...469..379E}
}

@ARTICLE{2020ApJ...893...46V,
       author = {{von Kienlin}, A. and {Meegan}, C.~A. and {Paciesas}, W.~S. and {Bhat}, P.~N. and {Bissaldi}, E. and {Briggs}, M.~S. and {Burns}, E. and {Cleveland}, W.~H. and {Gibby}, M.~H. and {Giles}, M.~M. and et al.},
        title = "{The Fourth Fermi-GBM Gamma-Ray Burst Catalog: A Decade of Data}",
      journal = {\apj},
         year = 2020,
        month = apr,
       volume = {893},
       number = {1},
          eid = {46},
        pages = {46},
          doi = {10.3847/1538-4357/ab7a18},
archivePrefix = {arXiv},
       eprint = {2002.11460},
 primaryClass = {astro-ph.HE},
       adsurl = {https://ui.adsabs.harvard.edu/abs/2020ApJ...893...46V}
}

@ARTICLE{2014ApJS..211...12G,
       author = {{Gruber}, David and {Goldstein}, Adam and {Weller von Ahlefeld}, Victoria and {Narayana Bhat}, P. and {Bissaldi}, Elisabetta and {Briggs}, Michael S. and {Byrne}, Dave and {Cleveland}, William H. and {Connaughton}, Valerie and {Diehl}, Roland and et al.},
        title = "{The Fermi GBM Gamma-Ray Burst Spectral Catalog: Four Years of Data}",
      journal = {\apjs},
         year = 2014,
        month = mar,
       volume = {211},
       number = {1},
          eid = {12},
        pages = {12},
          doi = {10.1088/0067-0049/211/1/12},
archivePrefix = {arXiv},
       eprint = {1401.5069},
 primaryClass = {astro-ph.HE},
       adsurl = {https://ui.adsabs.harvard.edu/abs/2014ApJS..211...12G}
}

@ARTICLE{2014ApJS..211...13V,
       author = {{von Kienlin}, Andreas and {Meegan}, Charles A. and {Paciesas}, William S. and {Bhat}, P.~N. and {Bissaldi}, Elisabetta and {Briggs}, Michael S. and {Burgess}, J. Michael and {Byrne}, David and {Chaplin}, Vandiver and {Cleveland}, William and et al.},
        title = "{The Second Fermi GBM Gamma-Ray Burst Catalog: The First Four Years}",
      journal = {\apjs},
         year = 2014,
        month = mar,
       volume = {211},
       number = {1},
          eid = {13},
        pages = {13},
          doi = {10.1088/0067-0049/211/1/13},
archivePrefix = {arXiv},
       eprint = {1401.5080},
 primaryClass = {astro-ph.HE},
       adsurl = {https://ui.adsabs.harvard.edu/abs/2014ApJS..211...13V}
}

@ARTICLE{2016ApJS..223...28N,
       author = {{Narayana Bhat}, P. and {Meegan}, Charles A. and {von Kienlin}, Andreas and {Paciesas}, William S. and {Briggs}, Michael S. and {Burgess}, J. Michael and {Burns}, Eric and {Chaplin}, Vandiver and {Cleveland}, William H. and {Collazzi}, Andrew C. and {Connaughton}, Valerie and {Diekmann}, Anne M. and {Fitzpatrick}, Gerard and {Gibby}, Melissa H. and {Giles}, Misty M. and {Goldstein}, Adam M. and {Greiner}, Jochen and {Jenke}, Peter A. and {Kippen}, R. Marc and {Kouveliotou}, Chryssa and {Mailyan}, Bagrat and {McBreen}, Sheila and {Pelassa}, Veronique and {Preece}, Robert D. and {Roberts}, Oliver J. and {Sparke}, Linda S. and {Stanbro}, Matthew and {Veres}, P{\'e}ter and {Wilson-Hodge}, Colleen A. and {Xiong}, Shaolin and {Younes}, George and {Yu}, Hoi-Fung and {Zhang}, Binbin},
        title = "{The Third Fermi GBM Gamma-Ray Burst Catalog: The First Six Years}",
      journal = {\apjs},
         year = 2016,
        month = apr,
       volume = {223},
       number = {2},
          eid = {28},
        pages = {28},
          doi = {10.3847/0067-0049/223/2/28},
archivePrefix = {arXiv},
       eprint = {1603.07612},
 primaryClass = {astro-ph.HE},
       adsurl = {https://ui.adsabs.harvard.edu/abs/2016ApJS..223...28N}
}

@ARTICLE{2006MNRAS.372..233A,
       author = {{Amati}, Lorenzo},
        title = "{The E$_{p,i}$-E$_{iso}$ correlation in gamma-ray bursts: updated observational status, re-analysis and main implications}",
      journal = {\mnras},
         year = 2006,
        month = oct,
       volume = {372},
       number = {1},
        pages = {233-245},
          doi = {10.1111/j.1365-2966.2006.10840.x},
archivePrefix = {arXiv},
       eprint = {astro-ph/0601553},
 primaryClass = {astro-ph},
       adsurl = {https://ui.adsabs.harvard.edu/abs/2006MNRAS.372..233A}
}

@ARTICLE{2023ApJ...946L..21N,
       author = {{Negro}, Michela and {Di Lalla}, Niccol{\`o} and {Omodei}, Nicola and {Veres}, P{\'e}ter and {Silvestri}, Stefano and {Manfreda}, Alberto and {Burns}, Eric and {Baldini}, Luca and {Costa}, Enrico and {Ehlert}, Steven R. and et al.},
        title = "{The IXPE View of GRB 221009A}",
      journal = {\apjl},
         year = 2023,
        month = mar,
       volume = {946},
       number = {1},
          eid = {L21},
        pages = {L21},
          doi = {10.3847/2041-8213/acba17},
archivePrefix = {arXiv},
       eprint = {2301.01798},
 primaryClass = {astro-ph.HE},
       adsurl = {https://ui.adsabs.harvard.edu/abs/2023ApJ...946L..21N}
}

@ARTICLE{2022GCN.32648....1D,
       author = {{de Ugarte Postigo}, A. and {Izzo}, L. and {Pugliese}, G. and {Xu}, D. and {Schneider}, B. and {Fynbo}, J.~P.~U. and {Tanvir}, N.~R. and {Malesani}, D.~B. and {Saccardi}, A. and {Kann}, D.~A. and {Wiersema}, K. and {Gompertz}, B.~P. and {Thoene}, C.~C. and {Levan}, A.~J. and {Stargate Collaboration}},
        title = "{GRB 221009A: Redshift from X-shooter/VLT}",
      journal = {GRB Coordinates Network},
         year = 2022,
        month = oct,
       volume = {32648},
        pages = {1},
       adsurl = {https://ui.adsabs.harvard.edu/abs/2022GCN.32648....1D}
}

@ARTICLE{2025SCPMA..6819505G,
       author = {{Ge}, Mingyu and {Ji}, Long and {Taverna}, Roberto and {Tsygankov}, Sergey and {Xu}, Yanjun and {Santangelo}, Andrea and {Zane}, Silvia and {Zhang}, Shuang-Nan and {Feng}, Hua and {Chen}, Wei and et al.},
        title = "{Physics of strong magnetism with eXTP}",
      journal = {Science China Physics, Mechanics, and Astronomy},
         year = 2025,
        month = sep,
       volume = {68},
       number = {11},
          eid = {119505},
        pages = {119505},
          doi = {10.1007/s11433-025-2796-y},
archivePrefix = {arXiv},
       eprint = {2506.08369},
 primaryClass = {astro-ph.HE},
       adsurl = {https://ui.adsabs.harvard.edu/abs/2025SCPMA..6819505G}
}

@ARTICLE{2025SCPMA..6819504B,
       author = {{Bu}, Qingcui and {Bambi}, Cosimo and {Gou}, Lijun and {Xu}, Yanjun and {Uttley}, Phil and {De Rosa}, Alessandra and {Santangelo}, Andrea and {Zane}, Silvia and {Feng}, Hua and {Zhang}, Shuang-Nan and et al.},
        title = "{Probing the strong gravity region of black holes with eXTP}",
      journal = {Science China Physics, Mechanics, and Astronomy},
         year = 2025,
        month = sep,
       volume = {68},
       number = {11},
          eid = {119504},
        pages = {119504},
          doi = {10.1007/s11433-025-2789-2},
archivePrefix = {arXiv},
       eprint = {2506.08105},
 primaryClass = {astro-ph.HE},
       adsurl = {https://ui.adsabs.harvard.edu/abs/2025SCPMA..6819504B}
}

@ARTICLE{2025SCPMA..6819503L,
       author = {{Li}, Ang and {Watts}, Anna L. and {Zhang}, Guobao and {Guillot}, Sebastien and {Xu}, Yanjun and {Santangelo}, Andrea and {Zane}, Silvia and {Feng}, Hua and {Zhang}, Shuang-Nan and {Ge}, Mingyu and et al.},
        title = "{Dense matter in neutron stars with eXTP}",
      journal = {Science China Physics, Mechanics, and Astronomy},
         year = 2025,
        month = sep,
       volume = {68},
       number = {11},
          eid = {119503},
        pages = {119503},
          doi = {10.1007/s11433-025-2761-4},
archivePrefix = {arXiv},
       eprint = {2506.08104},
 primaryClass = {astro-ph.HE},
       adsurl = {https://ui.adsabs.harvard.edu/abs/2025SCPMA..6819503L}
}

@ARTICLE{2022ApJ...934..109Q,
       author = {{Qi}, Liqiang and {Li}, Gang and {Ge}, Mingyu and {Zhang}, Juan and {Jiang}, Weichun and {Liu}, Xiaohua and {Yang}, Sheng and {Du}, Yuanyuan and {Dong}, Zefang and {Yang}, Yanji and et al.},
        title = "{Implementation of the Polarimetry Focusing Telescope Array Observation Simulator on board the X-Ray Timing and Polarimetry Observatory}",
      journal = {\apj},
         year = 2022,
        month = aug,
       volume = {934},
       number = {2},
          eid = {109},
        pages = {109},
          doi = {10.3847/1538-4357/ac7b82},
       adsurl = {https://ui.adsabs.harvard.edu/abs/2022ApJ...934..109Q}
}

@ARTICLE{2023ExA....56..517Q,
       author = {{Qi}, Liqiang and {Li}, Gang and {Xu}, Yupeng and {Zhang}, Juan and {Ge}, Mingyu and {Xiao}, Jingyu and {Ye}, Wentao and {Xiao}, Yunxiang and {Li}, Xiaobo},
        title = "{Application of the observation simulator in the eXTP mission}",
      journal = {Experimental Astronomy},
         year = 2023,
        month = aug,
       volume = {56},
       number = {2-3},
        pages = {517-536},
          doi = {10.1007/s10686-023-09910-y},
       adsurl = {https://ui.adsabs.harvard.edu/abs/2023ExA....56..517Q}
}

@ARTICLE{1959PhRvL...2...83W,
       author = {{Weibel}, Erich S.},
        title = "{Spontaneously Growing Transverse Waves in a Plasma Due to an Anisotropic Velocity Distribution}",
      journal = {\prl},
         year = 1959,
        month = feb,
       volume = {2},
       number = {3},
        pages = {83-84},
          doi = {10.1103/PhysRevLett.2.83},
       adsurl = {https://ui.adsabs.harvard.edu/abs/1959PhRvL...2...83W}
}

@ARTICLE{1999ApJ...526..697M,
       author = {{Medvedev}, Mikhail V. and {Loeb}, Abraham},
        title = "{Generation of Magnetic Fields in the Relativistic Shock of Gamma-Ray Burst Sources}",
      journal = {\apj},
         year = 1999,
        month = dec,
       volume = {526},
       number = {2},
        pages = {697-706},
          doi = {10.1086/308038},
archivePrefix = {arXiv},
       eprint = {astro-ph/9904363},
 primaryClass = {astro-ph},
       adsurl = {https://ui.adsabs.harvard.edu/abs/1999ApJ...526..697M}
}

@ARTICLE{2023ApJ...952L..42L,
       author = {{Lesage}, S. and {Veres}, P. and {Briggs}, M.~S. and {Goldstein}, A. and {Kocevski}, D. and {Burns}, E. and {Wilson-Hodge}, C.~A. and {Bhat}, P.~N. and {Huppenkothen}, D. and {Fryer}, C.~L. and {Hamburg}, R. and {Racusin}, J. and {Bissaldi}, E. and {Cleveland}, W.~H. and {Dalessi}, S. and {Fletcher}, C. and {Giles}, M.~M. and {Hristov}, B.~A. and {Hui}, C.~M. and {Mailyan}, B. and {Malacaria}, C. and {Poolakkil}, S. and {Roberts}, O.~J. and {von Kienlin}, A. and {Wood}, J. and {Ajello}, M. and {Arimoto}, M. and {Baldini}, L. and {Ballet}, J. and {Baring}, M.~G. and {Bastieri}, D. and {Gonzalez}, J. Becerra and {Bellazzini}, R. and {Bissaldi}, E. and {Blandford}, R.~D. and {Bonino}, R. and {Bruel}, P. and {Buson}, S. and {Cameron}, R.~A. and {Caputo}, R. and {Caraveo}, P.~A. and {Cavazzuti}, E. and {Chiaro}, G. and {Cibrario}, N. and {Ciprini}, S. and {Orestano}, P. Cristarella and {Crnogorcevic}, M. and {Cuoco}, A. and {Cutini}, S. and {D'Ammando}, F. and {De Gaetano}, S. and {Di Lalla}, N. and {Di Venere}, L. and {Dom{\'\i}nguez}, A. and {Fegan}, S.~J. and {Ferrara}, E.~C. and {Fleischhack}, H. and {Fukazawa}, Y. and {Funk}, S. and {Fusco}, P. and {Galanti}, G. and {Gammaldi}, V. and {Gargano}, F. and {Gasbarra}, C. and {Gasparrini}, D. and {Germani}, S. and {Giacchino}, F. and {Giglietto}, N. and {Gill}, R. and {Giroletti}, M. and {Granot}, J. and {Green}, D. and {Grenier}, I.~A. and {Guiriec}, S. and {Gustafsson}, M. and {Hays}, E. and {Hewitt}, J.~W. and {Horan}, D. and {Hou}, X. and {Kuss}, M. and {Latronico}, L. and {Laviron}, A. and {Lemoine-Goumard}, M. and {Li}, J. and {Liodakis}, I. and {Longo}, F. and {Loparco}, F. and {Lorusso}, L. and {Lovellette}, M.~N. and {Lubrano}, P. and {Maldera}, S. and {Manfreda}, A. and {Mart{\'\i}-Devesa}, G. and {Mazziotta}, M.~N. and {McEnery}, J.~E. and {Mereu}, I. and {Meyer}, M. and {Michelson}, P.~F. and {Mizuno}, T. and {Monzani}, M.~E. and {Morselli}, A. and {Moskalenko}, I.~V. and {Negro}, M. and {Nuss}, E. and {Omodei}, N. and {Orlando}, E. and {Ormes}, J.~F. and {Paneque}, D. and {Panzarini}, G. and {Persic}, M. and {Pesce-Rollins}, M. and {Pillera}, R. and {Piron}, F. and {Poon}, H. and {Porter}, T.~A. and {Principe}, G. and {Rain{\`o}}, S. and {Rando}, R. and {Rani}, B. and {Razzano}, M. and {Razzaque}, S. and {Reimer}, A. and {Reimer}, O. and {Ryde}, F. and {S{\'a}nchez-Conde}, M. and {Parkinson}, P.~M. Saz and {Scotton}, L. and {Serini}, D. and {Sgr{\`o}}, C. and {Sharma}, V. and {Siskind}, E.~J. and {Spandre}, G. and {Spinelli}, P. and {Tajima}, H. and {Torres}, D.~F. and {Valverde}, J. and {Venters}, T. and {Wadiasingh}, Z. and {Wood}, K. and {Zaharijas}, G.},
        title = "{Fermi-GBM Discovery of GRB 221009A: An Extraordinarily Bright GRB from Onset to Afterglow}",
      journal = {\apjl},
         year = 2023,
        month = aug,
       volume = {952},
       number = {2},
          eid = {L42},
        pages = {L42},
          doi = {10.3847/2041-8213/ace5b4},
archivePrefix = {arXiv},
       eprint = {2303.14172},
 primaryClass = {astro-ph.HE},
       adsurl = {https://ui.adsabs.harvard.edu/abs/2023ApJ...952L..42L}
}

@ARTICLE{2023ApJ...946L..24W,
       author = {{Williams}, Maia A. and {Kennea}, Jamie A. and {Dichiara}, S. and {Kobayashi}, Kohei and {Iwakiri}, Wataru B. and {Beardmore}, Andrew P. and {Evans}, P.~A. and {Heinz}, Sebastian and {Lien}, Amy and {Oates}, S.~R. and {Negoro}, Hitoshi and {Cenko}, S. Bradley and {Buisson}, Douglas J.~K. and {Hartmann}, Dieter H. and {Jaisawal}, Gaurava K. and {Kuin}, N.~P.~M. and {Lesage}, Stephen and {Page}, Kim L. and {Parsotan}, Tyler and {Pasham}, Dheeraj R. and {Sbarufatti}, B. and {Siegel}, Michael H. and {Sugita}, Satoshi and {Younes}, George and {Ambrosi}, Elena and {Arzoumanian}, Zaven and {Bernardini}, M.~G. and {Campana}, S. and {Capalbi}, Milvia and {Caputo}, Regina and {D'A{\`\i}}, Antonino and {D'Avanzo}, P. and {D'Elia}, V. and {De Pasquale}, Massimiliano and {Eyles-Ferris}, R.~A.~J. and {Ferrara}, Elizabeth and {Gendreau}, Keith C. and {Gropp}, Jeffrey D. and {Kawai}, Nobuyuki and {Klingler}, Noel and {Laha}, Sibasish and {Melandri}, A. and {Mihara}, Tatehiro and {Moss}, Michael and {O'Brien}, Paul and {Osborne}, Julian P. and {Palmer}, David M. and {Perri}, Matteo and {Serino}, Motoko and {Sonbas}, E. and {Stamatikos}, Michael and {Starling}, Rhaana and {Tagliaferri}, G. and {Tohuvavohu}, Aaron and {Zane}, Silvia and {Ziaeepour}, Houri},
        title = "{GRB 221009A: Discovery of an Exceptionally Rare Nearby and Energetic Gamma-Ray Burst}",
      journal = {\apjl},
         year = 2023,
        month = mar,
       volume = {946},
       number = {1},
          eid = {L24},
        pages = {L24},
          doi = {10.3847/2041-8213/acbcd1},
archivePrefix = {arXiv},
       eprint = {2302.03642},
 primaryClass = {astro-ph.HE},
       adsurl = {https://ui.adsabs.harvard.edu/abs/2023ApJ...946L..24W}
}

@ARTICLE{2024SCPMA..6789511Z,
       author = {{Zhang}, Yan-Qiu and {Xiong}, Shao-Lin and {Mao}, Ji-Rong and {Zhang}, Shuang-Nan and {Xue}, Wang-Chen and {Zheng}, Chao and {Liu}, Jia-Cong and {Zhang}, Zhen and {Wang}, Xi-Lu and {Ge}, Ming-Yu and et al.},
        title = "{Observation of spectral lines in the exceptional GRB 221009A}",
      journal = {Science China Physics, Mechanics, and Astronomy},
         year = 2024,
        month = aug,
       volume = {67},
       number = {8},
          eid = {289511},
        pages = {289511},
          doi = {10.1007/s11433-023-2381-0},
archivePrefix = {arXiv},
       eprint = {2403.12851},
 primaryClass = {astro-ph.HE},
       adsurl = {https://ui.adsabs.harvard.edu/abs/2024SCPMA..6789511Z}
}

@ARTICLE{2019ApJ...878L..26L,
       author = {{Laskar}, Tanmoy and {Alexander}, Kate D. and {Gill}, Ramandeep and {Granot}, Jonathan and {Berger}, Edo and {Mundell}, C.~G. and {Barniol Duran}, Rodolfo and {Bolmer}, J. and {Duffell}, Paul and {van Eerten}, Hendrik and et al.},
        title = "{ALMA Detection of a Linearly Polarized Reverse Shock in GRB 190114C}",
      journal = {\apjl},
         year = 2019,
        month = jun,
       volume = {878},
       number = {1},
          eid = {L26},
        pages = {L26},
          doi = {10.3847/2041-8213/ab2247},
archivePrefix = {arXiv},
       eprint = {1904.07261},
 primaryClass = {astro-ph.HE},
       adsurl = {https://ui.adsabs.harvard.edu/abs/2019ApJ...878L..26L}
}

@ARTICLE{2019ApJ...884..121L,
       author = {{Laskar}, Tanmoy and {van Eerten}, Hendrik and {Schady}, Patricia and {Mundell}, C.~G. and {Alexander}, Kate D. and {Barniol Duran}, Rodolfo and {Berger}, Edo and {Bolmer}, J. and {Chornock}, Ryan and {Coppejans}, Deanne L. and et al.},
        title = "{A Reverse Shock in GRB 181201A}",
      journal = {\apj},
         year = 2019,
        month = oct,
       volume = {884},
       number = {2},
          eid = {121},
        pages = {121},
          doi = {10.3847/1538-4357/ab40ce},
archivePrefix = {arXiv},
       eprint = {1907.13128},
 primaryClass = {astro-ph.HE},
       adsurl = {https://ui.adsabs.harvard.edu/abs/2019ApJ...884..121L}
}

@ARTICLE{2018ApJ...862...94L,
       author = {{Laskar}, Tanmoy and {Alexander}, Kate D. and {Berger}, Edo and {Guidorzi}, Cristiano and {Margutti}, Raffaella and {Fong}, Wen-fai and {Kilpatrick}, Charles D. and {Milne}, Peter and {Drout}, Maria R. and {Mundell}, C.~G. and et al.},
        title = "{First ALMA Light Curve Constrains Refreshed Reverse Shocks and Jet Magnetization in GRB 161219B}",
      journal = {\apj},
         year = 2018,
        month = aug,
       volume = {862},
       number = {2},
          eid = {94},
        pages = {94},
          doi = {10.3847/1538-4357/aacbcc},
archivePrefix = {arXiv},
       eprint = {1808.09476},
 primaryClass = {astro-ph.HE},
       adsurl = {https://ui.adsabs.harvard.edu/abs/2018ApJ...862...94L}
}

@ARTICLE{2018ApJ...859..134L,
       author = {{Laskar}, Tanmoy and {Berger}, Edo and {Margutti}, Raffaella and {Zauderer}, B. Ashley and {Williams}, Peter K.~G. and {Fong}, Wen-fai and {Sari}, Re'em and {Alexander}, Kate D. and {Kamble}, Atish},
        title = "{A VLA Study of High-redshift GRBs. II. The Complex Radio Afterglow of GRB 140304A: Shell Collisions and Two Reverse Shocks}",
      journal = {\apj},
         year = 2018,
        month = jun,
       volume = {859},
       number = {2},
          eid = {134},
        pages = {134},
          doi = {10.3847/1538-4357/aabfd8},
archivePrefix = {arXiv},
       eprint = {1707.05784},
 primaryClass = {astro-ph.HE},
       adsurl = {https://ui.adsabs.harvard.edu/abs/2018ApJ...859..134L}
}

@ARTICLE{2016ApJ...833...88L,
       author = {{Laskar}, Tanmoy and {Alexander}, Kate D. and {Berger}, Edo and {Fong}, Wen-fai and {Margutti}, Raffaella and {Shivvers}, Isaac and {Williams}, Peter K.~G. and {Kopa{\v{c}}}, Drejc and {Kobayashi}, Shiho and {Mundell}, Carole and et al.},
        title = "{A Reverse Shock in GRB 160509A}",
      journal = {\apj},
         year = 2016,
        month = dec,
       volume = {833},
       number = {1},
          eid = {88},
        pages = {88},
          doi = {10.3847/1538-4357/833/1/88},
archivePrefix = {arXiv},
       eprint = {1606.08873},
 primaryClass = {astro-ph.HE},
       adsurl = {https://ui.adsabs.harvard.edu/abs/2016ApJ...833...88L}
}

@ARTICLE{2013ApJ...776..119L,
       author = {{Laskar}, T. and {Berger}, E. and {Zauderer}, B.~A. and {Margutti}, R. and {Soderberg}, A.~M. and {Chakraborti}, S. and {Lunnan}, R. and {Chornock}, R. and {Chandra}, P. and {Ray}, A.},
        title = "{A Reverse Shock in GRB 130427A}",
      journal = {\apj},
         year = 2013,
        month = oct,
       volume = {776},
       number = {2},
          eid = {119},
        pages = {119},
          doi = {10.1088/0004-637X/776/2/119},
archivePrefix = {arXiv},
       eprint = {1305.2453},
 primaryClass = {astro-ph.HE},
       adsurl = {https://ui.adsabs.harvard.edu/abs/2013ApJ...776..119L}
}

@ARTICLE{2009Natur.462..767S,
       author = {{Steele}, I.~A. and {Mundell}, C.~G. and {Smith}, R.~J. and {Kobayashi}, S. and {Guidorzi}, C.},
        title = "{Ten per cent polarized optical emission from GRB090102}",
      journal = {\nat},
         year = 2009,
        month = dec,
       volume = {462},
       number = {7274},
        pages = {767-769},
          doi = {10.1038/nature08590},
archivePrefix = {arXiv},
       eprint = {1010.1255},
 primaryClass = {astro-ph.HE},
       adsurl = {https://ui.adsabs.harvard.edu/abs/2009Natur.462..767S}
}

@ARTICLE{2013Natur.504..119M,
       author = {{Mundell}, C.~G. and {Kopa{\v{c}}}, D. and {Arnold}, D.~M. and {Steele}, I.~A. and {Gomboc}, A. and {Kobayashi}, S. and {Harrison}, R.~M. and {Smith}, R.~J. and {Guidorzi}, C. and {Virgili}, F.~J. and et al.},
        title = "{Highly polarized light from stable ordered magnetic fields in GRB 120308A}",
      journal = {\nat},
         year = 2013,
        month = dec,
       volume = {504},
       number = {7478},
        pages = {119-121},
          doi = {10.1038/nature12814},
       adsurl = {https://ui.adsabs.harvard.edu/abs/2013Natur.504..119M}
}

@INCOLLECTION{2010xpnw.book..202L,
       author = {{Lazzati},, D.},
        title = "{X-ray polarization of gamma-ray bursts}",
    booktitle = {X-ray Polarimetry: A New Window in Astrophysics by Ronaldo Bellazzini},
         year = 2010,
       editor = {{Bellazzini}, Ronaldo and {Costa}, Enrico and {Matt}, Giorgio and {Tagliaferri}, Gianpiero},
        pages = {202},
          doi = {10.1017/CBO9780511750809.031},
       adsurl = {https://ui.adsabs.harvard.edu/abs/2010xpnw.book..202L}
}

@ARTICLE{2006ApJ...642..354Z,
       author = {{Zhang}, Bing and {Fan}, Y.~Z. and {Dyks}, Jaroslaw and {Kobayashi}, Shiho and {M{\'e}sz{\'a}ros}, Peter and {Burrows}, David N. and {Nousek}, John A. and {Gehrels}, Neil},
        title = "{Physical Processes Shaping Gamma-Ray Burst X-Ray Afterglow Light Curves: Theoretical Implications from the Swift X-Ray Telescope Observations}",
      journal = {\apj},
         year = 2006,
        month = may,
       volume = {642},
       number = {1},
        pages = {354-370},
          doi = {10.1086/500723},
archivePrefix = {arXiv},
       eprint = {astro-ph/0508321},
 primaryClass = {astro-ph},
       adsurl = {https://ui.adsabs.harvard.edu/abs/2006ApJ...642..354Z}
}

@ARTICLE{2006ApJ...642..389N,
       author = {{Nousek}, J.~A. and {Kouveliotou}, C. and {Grupe}, D. and {Page}, K.~L. and {Granot}, J. and {Ramirez-Ruiz}, E. and {Patel}, S.~K. and {Burrows}, D.~N. and {Mangano}, V. and {Barthelmy}, S. and et al.},
        title = "{Evidence for a Canonical Gamma-Ray Burst Afterglow Light Curve in the Swift XRT Data}",
      journal = {\apj},
         year = 2006,
        month = may,
       volume = {642},
       number = {1},
        pages = {389-400},
          doi = {10.1086/500724},
archivePrefix = {arXiv},
       eprint = {astro-ph/0508332},
 primaryClass = {astro-ph},
       adsurl = {https://ui.adsabs.harvard.edu/abs/2006ApJ...642..389N}
}

@ARTICLE{2007ApJ...662.1093W,
       author = {{Willingale}, R. and {O'Brien}, P.~T. and {Osborne}, J.~P. and {Godet}, O. and {Page}, K.~L. and {Goad}, M.~R. and {Burrows}, D.~N. and {Zhang}, B. and {Rol}, E. and {Gehrels}, N. and et al.},
        title = "{Testing the Standard Fireball Model of Gamma-Ray Bursts Using Late X-Ray Afterglows Measured by Swift}",
      journal = {\apj},
         year = 2007,
        month = jun,
       volume = {662},
       number = {2},
        pages = {1093-1110},
          doi = {10.1086/517989},
archivePrefix = {arXiv},
       eprint = {astro-ph/0612031},
 primaryClass = {astro-ph},
       adsurl = {https://ui.adsabs.harvard.edu/abs/2007ApJ...662.1093W}
}
\end{document}